\documentclass[letterpaper]{CVIS}
\usepackage{graphicx}% Include figure files
\usepackage{dcolumn}% Align table columns on decimal point
\usepackage{subcaption,soul}
\usepackage{caption}
\usepackage[authoryear]{natbib}
\usepackage{url}
\usepackage{mleftright}
\usepackage{multirow}
\usepackage{titlesec}
\usepackage[redeflists]{IEEEtrantools}
\usepackage[pagebackref=false,breaklinks=true,colorlinks=true,linkcolor = blue, urlcolor = black,citecolor=black,bookmarks=true]{hyperref}

\usepackage{amssymb}
\newgeometry{left=0.6in,right=0.6in,top=1in,bottom=0.9in}
\usepackage[utf8]{inputenc} 
\usepackage[T1]{fontenc}
\usepackage[british]{babel}
\usepackage[
  input-uncertainty-signs=\pm,
  separate-uncertainty = true,
  group-separator = {\,},
	detect-weight = true,
  detect-family = true,
  power-half-as-sqrt = true,
]{siunitx}
\DeclareSIUnit[]\km{\kilo\meter}
\DeclareSIUnit[]\mumeter{\micro\meter}
\DeclareSIUnit[]\um{\micro\meter}
\DeclareSIUnit[]\uHz{\micro\hertz}
\DeclareSIUnit[]\mHz{\milli\hertz}
\DeclareSIUnit[per-mode=symbol]\nms{\nm\per\s\squared}
\DeclareSIUnit[per-mode=symbol]\mps{\m\per\s\squared}

\usepackage[version=4]{mhchem} 
\usepackage[]{graphicx} % Einbindung von Grafiken (pdf, png, jpg)
\usepackage{amsmath}
\usepackage{bm}
\usepackage{xspace}
\usepackage{hyperref}
\usepackage{tablefootnote}
\definecolor{newcolor}{rgb}{.8,.349,.1}
\newcommand{\eg}{e.\,g.\@\xspace}

\newcommand{\msqwert}[1]{\qty{#1}{\mps\per\sqrt{\text{\Hz}}}}
\newcommand{\vkeff}{\vec{k}_\ensuremath{\mathrm{eff}}\xspace}
\newcommand{\keff}{k_\ensuremath{\mathrm{eff}}\xspace}
\renewcommand{\vec}[1]{\ensuremath{\mathbf{#1}}}

\title{Advances in Atom Interferometry and their Impacts on the Performance of Quantum Accelerometers On-board Future Satellite Gravity Missions}

\begin{document}
%\bstctlcite{IEEEexample:BSTcontrol}
%\sloppy
% Please enter author Names, Affiliations, and Emails. 
% Use superscripts to indicate the affiliation, which is then specified in a row beneath all the names. 
% Author names should be separated by a \quad space.
% If a second line of Author names is required, each name in the second row must be manually bolded via \textbf{Author name}.
% The \thanks{} command may be used to provide a footnote indicating additional information for each author.
\author{
Alireza HosseiniArani$^{1,*}$\quad Manuel Schilling$^{2}$ \quad Quentin Beaufils$^{3}$ \quad Annike Knabe$^{1}$  \quad Benjamin Tennstedt$^{1}$\\  \quad \textbf{Alexey Kupriyanov}$^{1}$ \quad \textbf{Steffen Schön}$^{1}$ \quad \textbf{Franck Pereira dos Santos}$^{3}$ \quad \textbf{Jürgen Müller}$^{1}$\\ 
$^1$Institute of Geodesy (IfE), Leibniz University Hannover\\
$^2$Institute for Satellite Geodesy and Inertial Sensing, German Aerospace Center (DLR)\\
$^3$LNE-SYRTE, Observatoire de Paris, Université PSL, CNRS, Sorbonne Université\\
\texttt{\ * hosseiniarani@ife.uni-hannover.de}\\
}

\maketitle

% \begin{abstract} % maximum of 150 words
% 	This template presents the author guidelines for all submissions for Journal of Computational Vision and Imaging Systems. Please ensure that all guidelines are followed prior to submission.
% \end{abstract}

\begin{abstract}
%%%
Recent advances in cold atom interferometry have cleared the path for space applications of quantum inertial sensors, whose level of stability is expected to increase dramatically with the longer interrogation times accessible in space. In this study, a comprehensive in-orbit model is developed for a Mach-Zehnder-type cold-atom accelerometer. Performance tests are realized under different assumptions, and the impact of various sources of errors on instrument stability is evaluated. Current and future advances for space-based atom interferometry are discussed, and their impact on the performance of quantum sensors on-board satellite gravity missions is investigated in three different scenarios: state-of-the-art scenario, near-future (between the next 5 and 10 years) and far-future scenarios (between the next 10 to 20 years). We show that one can achieve a sensitivity level close to \msqwert{5e-10} with the current state-of-the-art technology. We also estimate that in the near and far-future, atom interferometry in space is expected to achieve sensitivity levels of \msqwert{1e-11} and \msqwert{1e-12}, respectively. A roadmap for improvements in atom interferometry is provided that would maximize the performance of future CAI accelerometers, considering their technical capabilities. Finally, the possibility and challenges of having ultra-sensitive atom interferometry in space for future space missions are discussed. 

%%%%
\end{abstract}

% \section{Introduction}

% \begin{keyword}
% %% MSC codes here, in the form: \MSC code \sep code
% %% or \MSC[2008] code \sep code (2000 is the default)
% %\MSC 41A05\sep 41A10\sep 65D05includes the envoironemtal noise \sep 65D17
% %% Keywords
% \KWD Atom interferometry\sep  CAI accelerometer\sep Quantum sensors\sep Inertial sensors\sep Satellite gravimetry\sep Satellite gravity missions 
% \end{keyword}

% \end{frontmatter}

%% For linenumbers
%\linenumbers

%% main text
\section{Introduction}\label{sec:Intro}

%Please use \verb+elsarticle.cls+ for typesetting your paper. Additionally,
%make sure not to remove the package \verb+jasr.sty+ already included in the
%preamble:
%\begin{verbatim} 

%\end{verbatim}

%Make sure to have the file \verb+jasr-model5-names.bst+ to produce the references in
%the correct format. 
%Any instructions relevant to the \verb+elsarticle.cls+ are
%applicable here as well. See the online instruction available at:
%\makeatletter
%\if@twocolumn
%\begin{verbatim}
% https://support.stmdocs.in/wiki/
% index.php?title=Elsarticle.cls
%\end{verbatim}
%\else
%\begin{verbatim}
%% https://support.stmdocs.in/wiki/index.php?title=Elsarticle.cls
% \end{verbatim}
%\fi
%\makeatother
       
%Following commands are defined for this journal which are not in
%\verb+elsarticle.cls+. 
%\begin{verbatim}
% \received{}
%  \finalform{}
%  \accepted{}
%  \availableonline{}
%  \communicated{}
%\end{verbatim}

%\subsection{Subsection}
%This is only a \LaTeX\ template if you need one. 
%See the detailed guidelines for manuscript preparation 
%and submission at:
%\begin{verbatim}
%https://ees.elsevier.com/asr
%\end{verbatim}

\subsection{Satellite Gravity Missions}\label{gravityMissions}

Satellite gravimetry missions monitor the Earth's gravity field and its changes over time. Results from previous missions like GRACE(-FO) contributed to quantifying mass variations related to climate change \citet{Tapley2019,Humphrey2023,Scanlon2023} and brought new insights into processes of the Earth’s interior \citet{Mandea2020, Lecomte2023}. But current solutions of the gravity field provided by these satellite gravity missions are limited in the very low degrees at $C_{20}$ and, for times with only one operational accelerometer on two satellites, even at $C_{30}$ \citet{Loomis2020}. These coefficients are thus typically replaced with satellite laser-ranging solutions. %The drift of the used electrostatic accelerometer at lower frequencies limits the solution. 
The drift in low frequencies of the electrostatic accelerometers used limits the gravity field solution. 
Overall, the spatial resolution is limited to $>(\qty{400}{\km})^2$ for a signal amplitude of \qty{10}{\milli\m} for typical monthly gravity field solutions. 
To address the needs of the scientific community, future satellite gravity missions shall target a resolution of $(\qty{200}{\km})^2$ for an amplitude of \qty{10}{\milli\m} equivalent water height or even smaller \citet{Pail2015, Wiese2022}.
%a resolution at the order of 100 km or below. 
This would allow, among others, more accurate measurements of the mass balance of smaller glaciers or sea level changes, as well as drought or flood predictions on a regional scale.
Potential technologies to overcome current limitations are, for example, improvements of electrostatic accelerometers by enhanced readout schemes \citet{Alvarez2022, Kupriyanov2024}, sensors based on atom interferometry \citet{Leveque2009} and combinations of classical and quantum accelerometers in a hybrid configuration \citet{Zahzam2022}. This work focuses on cold atom interferometry and hybrid concepts.

\subsection{Cold Atom Interferometry}\label{CAI}

%Since it allows for the realization of high-performance inertial sensors \cite{Geiger2020}, atom interferometry is a promising candidate technology for performing accelerometry in satellites. 
Atom interferometry is a promising candidate technology for performing accelerometry in satellites as it allows the realisation of high-performance inertial sensors with a flat noise spectrum \citet{Geiger2020}. 
In such sensors, atoms in free fall are used as test masses. Their acceleration in the satellite frame is precisely measured by realizing an atom interferometer with sequences of laser pulses. 
Such light-pulse atom interferometers can be implemented in a number of different ways \citet[see ][for an overview]{Abend2020}, and only a brief introduction based on \cite{Kasevich1991} is given here. 
The atom interferometer is implemented by three laser pulses acting either as beam splitters or mirror pulses. These laser pulses consist of two counter-propagating laser beams, whose frequency difference is tuned in resonance with a two-photon Raman transition between the two hyperfine ground states of an alkali atom, \ce{^{87}Rb} in our case. 
The first pulse of light acts as a beam splitter for matter waves. It places the atom in a quantum superposition of two wave packets of different momenta that spatially separate after the pulse during a time interval $T$. A second light pulse inverts the momenta of the two wave packets. Finally, a beamsplitting pulse closes the interferometer after a further time interval $T$.
The atom interferometer phase shift $\Delta\Phi$ depends on the projection of the acceleration $\vec{a}$ experienced by the atoms along the effective optical wave vector of the laser light $\vkeff=\vec{k}_1-\vec{k}_2$ which is the difference between the optical wave vectors $\vec{k}_i$ of both laser beams.
The leading order of the atom interferometer phase $\Delta\Phi$ is described by
\begin{equation}    \label{eq:CAIphase}
    \Delta\Phi= (2 ~ \keff ~ a)  T^2 + \Phi_L
\end{equation}
with the acceleration $a$ and $\keff=|\vkeff|$ now expressed in the direction of the counterpropagating laser beams. 
$\keff$ is also related to the photon momentum transfer induced by the Raman transitions. 

When the interferometer is operated in single diffraction, an arbitrary Raman laser phase $\Phi_L$ can be added to the last light pulse to scan the fringe pattern, or to operate the interferometer, \eg, at mid-fringe as described in~\cite{hosseiniarani2022}. In double diffraction \citet{Leveque2009}, as anticipated for in microgravity, this can be achieved by shifting the reference mirror's position, \eg with a piezoelectric actuator.  

%The linear frequency ramp $\alpha$ is added to compensate for the acceleration of the atoms. 
%The acceleration in direction of $\vkeff$ is canceled out if $\alpha=|\vkeff\cdot\vec{a}|$.
%This way, all measurements of a CAI can be traced back to (laser) frequencies.
%More specifically, $\Delta\Phi$ measures the projection of the acceleration $\vec{a}$ along $\vec{k_{eff}}$. In terrestrial applications, $\vec{k_{eff}}$ can be aligned with $\vec{g}$ to create a gravimeter. 
In a satellite setting, the instrument can be used to measure the non-gravitational accelerations acting on the satellite. A three-axis instrument could also be implemented for geodesy applications but in this paper we focus on a single-axis quantum accelerometer oriented in the along-track (X-) direction. Eq.~\eqref{eq:CAIphase} describes a single-axis accelerometer phase shift in the absence of rotation of the satellite. A more complete version of this equation including the first-order effects of rotations is shown in Eq.~\eqref{eq:CAIphasePlusRot} (section~\ref{sec: rotModel}). 

Currently, the technology to operate a quantum accelerometer in space is still under development. 
While quantum gravimeters are available for terrestrial applications, including commercial instruments, space applications are in a much more experimental state. 
Experiments on sounding rockets demonstrated the generation of a Bose-Einstein Condensate in space with the MAIUS Experiment \citet{Becker2018}. 
The follow up missions MAIUS-2/3 are planned to perform differential acceleration measurements between two species of atoms to test the Einstein equivalence principle \citet{Elsen2023}. 
Additionally, NASA operates the Cold Atom Lab (CAL) onboard the International Space Station \citet{Aveline2020} since 2018 which will be superseded by the Bose Einstein Condensate and Cold Atom Laboratory \citet[BECCAL; ][]{Frye2021} in the near future.
The aforementioned experiments do not fulfill the requirements of, \eg, volume or power consumption of a satellite platform which is typically used for a gravimetry mission.
The Horizon Europe funded CARIOQA Pathfinder Mission Preparation project\footnote{\url{https://doi.org/10.3030/101081775  }} aims to build a quantum accelerometer engineering model.
This project has been supplemented in 2024 by the CARIOQA Phase A study\footnote{\url{https://cordis.europa.eu/project/id/101135075}} in order to deploy a quantum space gravimetry pathfinder mission \citet{Leveque2022} before the year 2030.

\subsection{Quantum Accelerometers On-board Future Gravimetry Missions}\label{sec:introFutureCAI}

According to Eq.~\eqref{eq:CAIphase}, the sensitivity of the CAI can be increased by increasing the interrogation time $T$. In terrestrial applications, $T$ is limited by the length of the free fall distance of the atoms, \eg up to \SI{300}{\milli\s} for a transportable \citet{Freier2016} and up to a couple of seconds for stationary instruments \citet{Asenbaum2020,Schilling2020}. 
As atoms and satellites in space are in free fall, longer separation times $T$ are possible. There, a quantum accelerometer would allow for monitoring the deviation from the free fall trajectory resulting from non-gravitational accelerations acting on the satellite. A limiting factor on the maximum achievable interrogation time is then the residual thermal expansion of the atomic cloud. 

%Considering the intended duration of a single interferometer sequence, using Bose-Einstein Condensates might also be the preferred choice \citet{Becker2018}. 

%It is shown that using CAI accelerometers could be beneficial for future satellite gravity missions (\cite ?) and would allow quantifying the mass changes related to climate change \citet{Tapley2019} and new insights into processes of the Earth's interior \citet{Mandea2020}. 

The potential sensitivity gain allowed by the increase of the interrogation time up to a few seconds would make CAI accelerometers a competitive technology for future satellite gravity missions \citet{Abrykosov2019}. They indeed provide absolute measurements and high long-term stability. While this would also be highly beneficial in a hybridized configuration with classical sensors (\eg a relative electrostatic accelerometer (E-ACC)), the present study focuses on the performance of the CAI itself.

%Our assumption in this study is that we are benefiting from a hybrid accelerometer configuration, which means that in parallel with the CAI accelerometer, there is an electrostatic accelerometer (E-ACC) available which recovers the high-frequency variation of non-gravitational acceleration and allows us to solve for the CAI ambiguity. Details of the hybridization and performance will be the subject of another study. Here, we are only interested in the performance of the CAI accelerometer itself. 

\cite{hosseiniarani2022} have shown a Kalman-filter-based hybridization strategy of an electrostatic accelerometer with the characteristics of the GRACE-FO mission with a CAI accelerometer based on state-of-the-art technology to create a hybrid accelerometer on-board a future satellite gravity mission. \cite{hosseiniarani2022} assumes an achievable sensitivity of \msqwert{1e-10} based on an improvement of two order of magnitudes from ground applications in gravimetry \citet{Merlet2021} thanks to an extension of the interrogation time to several seconds. 
Additionally, the objective of current technology developments is also at the \msqwert{e-10} level, \eg in the Horizon Europe funded CARIOQA-PMP project \citet{Leveque2022} preparing a demonstration of a quantum accelerometer in space.
%Future gravity missions, however, will benefit from several advances in atom interferometry, e.g., lower atomic temperatures, and a higher laser waist. As a consequence, certain improvements in the noise levels are anticipated, which would allow the determination of the static and temporal gravity fields with higher sensitivity. 

In this paper, we will investigate the current and future advances in atom interferometry and study their impacts on the performance of quantum sensors on-board future satellite gravity missions. 
In section \ref{sec:modeling}, we describe our modeling environment and the theoretical background of the cold atom interferometer model. In section \ref{sec:sensi}, we discuss the effect of various parameters on the atom interferometer sensitivity. Finally, in section \ref{sec:orbitperf} we evaluate different atom interferometer configurations for satellite applications.

\section{Modeling} \label{sec:modeling}

\subsection{Orbit Model}
%\begin{verbatim} 
  %\end{verbatim}
We consider a GRACE-like satellite pair in a circular polar orbit around the Earth with an altitude of 480 km. The simulation is implemented in the MATLAB/Simulink-based eXtended High-Performance satellite dynamics Simulator \citet[XHPS; ][]{Woeske2019} developed by ZARM/DLR. XHPS calculates the orbits of a GRACE-FO mission scenario under consideration of the Earth’s gravity field \citet[EGM 2008 up to d/o 90; ][]{Pavlis2012}, non-gravitational forces (atmospheric drag, solar radiation pressure, Earth albedo and thermal radiation pressure) and the GRACE-FO satellite geometry. To consider the effect of non-gravitational forces on the spacecraft, we use a detailed surface model of the satellite body included in XHPS.

\subsection{CAI Accelerometer Signal Model} \label{signalModel}

%Each measurement of the CAI accelerometer has instrumental noise and quantum noise, while they are also affected by external effects like rotations. In this study, we consider all these effects. 

%This paper considers several noise sources such as quantum projection noise, technical 

The highest contribution to the signal of the CAI accelerometer comes from the non-gravitational accelerations. The phase shift generated by a constant acceleration signal would be given by Eq.~\eqref{eq:CAIphase}. However, since the non-gravitational acceleration varies during the CAI interrogation time, we use the integration form of Eq.~\eqref{eq:CAIphase} by considering the sensitivity function as described in \cite{Knabe2022}. The phase of interferometer $\Phi_k$ at the k-th cycle is given by

\begin{equation}    \label{eq:CAIphaseInt}
    \Delta\Phi= 2~\keff\left[ \int_{k~T_c}^{(k+1)~T_c} g_{a,k}~ a(t) \,dx \right]
\end{equation}

where $g_{a,k}$ is the sensitivity of the instrument. It is maximum at the mid-fridge and minimum at the beginning and end of the interferometry cycle.

\subsection{Modeling of the Rotational Effects} \label{sec: rotModel}

In addition to the phase shift caused by the non-gravitational signal, we consider rotational contributions to the interferometer phase, which come from the fact that the satellite rotates about its cross-track (Y-) axis with a rotation rate of \qty[per-mode=symbol]{\approx 1.1}{\milli\radian\per\s} to stay in a nadir-pointing orientation. This rotation creates additional phase contributions, which depend on the position of the CAI accelerometer inside the satellite and also the direction of its sensitivity axis.

The largest contribution of this rotation to the phase shift of the atom interferometer arises from the Coriolis acceleration induced by the atomic velocity in the radial direction \citet{Leveque2021}. In addition, there are also Euler and centrifugal contributions. 

%The positioning of the CAI accelerometer inside the satellite can play an important role in the achievable sensitivity of the CAI accelerometer. Figure~\ref{fig:CAIpos} compares the different possible positions for the CAI accelerometer inside the satellite. Here, the assumption is that because of the positioning of E-ACC at the satellite center of mass, the CAI accelerometer should be necessarily positioned somewhere else. Three possibilities are compared. Figure~\ref{fig:CAIpos}-a shows a possible case where the CAI accelerometer is placed in front of the E-ACC on the along-track axis. Figure~\ref{fig:CAIpos}-b shows another possibility in which the CAI accelerometer is positioned on top of the E-ACC and on the radial axis. Finally, Figure~\ref{fig:CAIpos}-c shows the case in w in which the CAI accelerometer is placed beside the E-ACC and on the cross-track axis of the satellite. 

As discussed in section~\ref{sec:introFutureCAI}, in this study we consider the CAI accelerometer to be used in combination with E-ACC. Since both instruments cannot be co-located, placing the E-ACC at the center of mass of the satellite requires a displacement of the position of the atoms with respect to the center of mass, resulting in a gravity gradient and a gravitational pull of the spacecraft's mass on the atoms.

The positioning of the CAI accelerometer within the satellite can play an important role in the achievable sensitivity. Figure~\ref{fig:CAIpos} compares the different possible positions for the CAI accelerometer inside the satellite. Here, the E-ACC is assumed to be at the center of mass of the satellite while the CAI is shifted from it. The CAI accelerometer is placed either in front of the E-ACC on the along-track axis, on top of the E-ACC on the radial axis or next to the E-ACC on the cross-track axis of the satellite. 

In a configuration similar to Fig.~\ref{fig:CAIpos}-a, the centrifugal and Coriolis accelerations are in the direction of the sensitivity axis, but the Euler acceleration is perpendicular and, therefore, is not sensed by the CAI accelerometer. In a configuration similar to Fig.~\ref{fig:CAIpos}-b, the Coriolis and Euler accelerations are aligned with the CAI accelerometer sensitivity axis, and the centrifugal acceleration is perpendicular to it. In configuration b, because of the displacement in the radial direction, the Earth's gravity gradient imposes an additional error on the CAI accelerometer. In the third configuration~(Fig.~\ref{fig:CAIpos}-c), the centrifugal and Coriolis accelerations are aligned with the sensitivity axis. However, the centrifugal acceleration has a considerably smaller value (close to zero) because the lever arm is now reduced to a distance close to the movement of the center of mass. As a consequence, the latest positioning shows considerable advantages compared to the other two possible cases. 

This phase shift due to the satellite rotation can be calculated from the following equations \citet{Beaufils2023}: 
\begin{eqnarray}    \label{eq:CAIphasePlusRot}
%%\begin{split}
    & \Delta\Phi= 2 \keff T^2 [a_x + 2 v_{z0} (\Omega_y + \Omega_M) - x_0 {\Omega_y}^2 \\
    & + (x_0 - x_M)({\Omega_M}^2 + {(\Omega_M - \Omega_I)}^2)]
%%\end{split}
\end{eqnarray}

where $x_0$ is the initial distance of atoms to the satellite center of mass, $x_M$ is the distance of the center of rotation of the mirror to the satellite center of mass, $\Omega_y$ is the angular velocity of the satellite around the cross-track axis with respect to the inertial frame, $\Omega_M$ and $\Omega_I$ are the angular velocities of the mirror and incoming laser beam, with respect to the satellite body-fixed frame, and $v_{z0}$ is the initial velocity of atoms in the radial direction and in the satellite frame. 

The rotational phase shift could add extremely large errors to the measurements of the quantum sensor if they are not properly compensated. Different approaches have been proposed to compensate for the rotation effect. The impact of the main rotation due to the orbital frequency can be compensated by counter-rotating the Raman wave vector with a fixed rate; \citet[see \eg, ][for a nadir pointing gradiometer]{Trimeche2019}. This approach would not compensate for the residual rotation error. 

In another approach, a high-performance onboard gyroscope can be used to measure the satellite rotation at each instant in time and cancel its contribution to the phase shift by an active Raman mirror rotating against the rotation rate of the satellite \citet{Lan2012,Migliaccio2019}. In this scenario, the incoming laser is assumed to be fixed in the body frame of the satellite and has a rotation rate identical to that of the satellite, therefore 
\begin{eqnarray}    \label{eq:mirRot}
%\begin{split}
     \Omega_M &= -\Omega_y \quad\text{and}\\
     \Omega_I &= 0.
%\end{split}
\end{eqnarray}
We also assume that the center of rotation of the mirror is the same as the center of mass of the satellite ($x_M = 0$). With these assumptions, Eq.~\eqref{eq:CAIphasePlusRot} becomes
\begin{eqnarray}    \label{eq:mirRot2}
  \Delta\Phi= 2 \keff T^2 [a_x + x_0 {\Omega_y}^2].
\end{eqnarray}
  %+ 2 v_y \Omega_z - x_0 {\Omega_z}^2 + z_0 \Omega_x \Omega_z +  y_0 \Omega_y \Omega_x]
The term $x_0 {\Omega_y}^2$ would be the remaining term which could induce a bias in the measurements of the CAI accelerometer. We need to stress that apart from the remaining term, the phase shift contains residual terms related to the difference between the rotation rate of the mirror and the satellite rotation rate due to imperfect calibration and noise in the rotation sensors and actuators. 
Furthermore, there are smaller rotation components around the other axes in addition to the main rotation of the satellite around the cross-track axis.
These residual terms would also result in additional phase shift components. For a comprehensive and accurate modeling, these residual terms are also accounted for in this study. 

The third approach considered in this study is the counter-rotation of the entire quantum sensor. This approach would be technically challenging. Nevertheless, we investigated it and compared the performance of the quantum sensor when different rotation compensation techniques were applied. The assumption for this approach is:

\begin{eqnarray}    \label{eq:WholeCAIrot}
     \Omega_M = \Omega_I = - \Omega_y .
\end{eqnarray}
With this assumption, Eq.~\eqref{eq:CAIphasePlusRot} will become
\begin{eqnarray}    \label{eq:WholeCAIrot2}
  \Delta\Phi= 2 \keff T^2 [a_x - x_M {\Omega_M}^2],
\end{eqnarray}
which means that, if the center of rotation of the Raman mirror is the same as the satellite center of mass ($x_M = 0$), there would be no remaining term. Please note that similar to the previous case, the phase shift contains residual terms related to the difference between the rotation rate of the mirror and the satellite rotation rate due to imperfect calibration and noise in the rotation sensors and actuators. However, these residual terms are negligible.

% the error due to the residual terms will have to be considered in the phase shift, but there would not be any remaining term which causes any bias on the measurements of the CAI accelerometer. 

\subsection{CAI Accelerometer Noise Model} \label{noiseModel}
In this study, we consider a full noise model that includes all the major noise sources that affect the measured phase shift. In section~\ref{sec: rotModel}, we discussed the environmental noises that impact the CAI accelerometer measurements. Below, we discuss the quantum and instrumental errors that are modelled in this study:  
%\paragraph{test headline}
%Could we maybe use the \emph{paragraph} environment instead of \emph{itemize} to list the following things? It would look like this paragraph.

%\begin{itemize}

\paragraph{Frequency noise of the master Raman laser ($\sigma_{\phi}$)}

While the noise in the phase difference between the lasers does not impact the sensitivity of the measurement in the double diffraction mode \citet{Leveque2009}, the frequency noise of the master Raman laser does \citet{legouet2007}. We assume here a white noise for the master Raman laser frequency. This noise contribution then scales inversely to the duration of the Raman pulses and proportionally to the distance to the mirror \citet{legouet2007}. 

\paragraph{Wavefront aberrations noise}    
 
Wavefront aberrations induce noise in the interferometer phase \citet{Louchet-Chauvet2011} that originates from the motion of the atoms in the distorted phase profile of the Raman laser beams. Generally speaking, this noise gets lower with a flatter Raman mirror, with a lower atomic temperature, and with a shorter interrogation time. It fluctuates with the initial position and velocity of the atomic source with respect to the laser beam, adding noise to the measurement.

\paragraph{Detection noise ($\sigma_p $)}
Detection noise has two components: Quantum projection noise (QPN), which is the quantum standard limit in quantum inertial sensors and clocks, and technical noise (TN), which is an electronic noise on the measurement of the number of atoms. Both QPN and TN get lower with a higher number of atoms. TN also has a constant term, which is a contribution independent of the number of atoms. So one has

\begin{eqnarray}    \label{eq:QPN}
     {\sigma_P}^2(QPN) = \dfrac{P(1-P)}{N},
\end{eqnarray}
\begin{eqnarray}    \label{eq:TN}
     {\sigma_P}^2(TN) = \dfrac{P^2 {\sigma_{N_2}}^2 + {(1-P)}^2 {\sigma_{N_1}}^2}{N^2} + {\sigma_P}^2(\infty),
\end{eqnarray}
\begin{eqnarray}    \label{eq:DetNoise}
     {\sigma_P}^2(DET) = {\sigma_P}^2(QPN) + {\sigma_P}^2(TN).
\end{eqnarray}

where, ${\sigma_P}^2(QPN)$ and ${\sigma_P}^2(TN)$ and ${\sigma_P}^2(DET)$ are the variances of the transition probability due to the QPN, TN and total detection noise respectively $\sigma_{N_i}$ is the electronic noise on the measurement of the number of atoms in the i-th port and ${\sigma_P}^2(\infty)$ is a contribution independent of the number of atoms. 
The latest is the dominating factor at a very large number of atoms. 
It is related to the frequency/intensity noise of the detection laser and to the normalisation noise.

\paragraph{Contrast loss}
Loss of contrast in the atom interferometer results in a loss of sensitivity in the measurement of the phase shift. It has two main sources. The first is the inhomogeneity of the laser intensity experienced by the atoms. Since the atomic cloud has a finite size and residual expansion, laser intensity inhomogeneity across the atomic cloud leads to coupling inhomogeneities, which induce losses in the number of atoms and in contrast. The second source is Coriolis acceleration, which leads to inhomogeneous dephasing due to the finite temperature of the atomic cloud. Averaging the Coriolis acceleration over the velocity distribution then leads to a loss of contrast. 

The loss of contrast due to the laser intensity inhomogeneity gets lower with lower atomic temperatures and shorter interrogation times. It also gets lower with higher laser waists. The loss due to Coriolis acceleration gets lower with lower angular velocities, shorter interrogation times, and lower temperatures ($T_0$).

%\paragraph{Transition probability noise}

%All the above-listed noise contributions lead to noise in the actual measurand, which is the transition probability at the end of the interferometry cycle. 

%\end{itemize}

%\subsection{Advances in atom intererometry}\label{subsecOngoing}

\section{Sensitivity Analysis}\label{sec:sensi}

\subsection{Sensitivity to the Positioning and Rotation}\label{rotComp}

It will be shown later that the rotational phase shift can wash out the contrast if not properly compensated (see Fig.~\ref{fig:stateArt1}). In section \ref{sec: rotModel}, we discussed three possible approaches to physically compensate for the rotation of the satellite. In this section, we discuss the impact of the rotation compensation method used together with the effects of the positioning of the CAI accelerometer inside the satellite frame, on the measurement noise.

\begin{figure*}[htbp]%
\includegraphics[width=\textwidth]{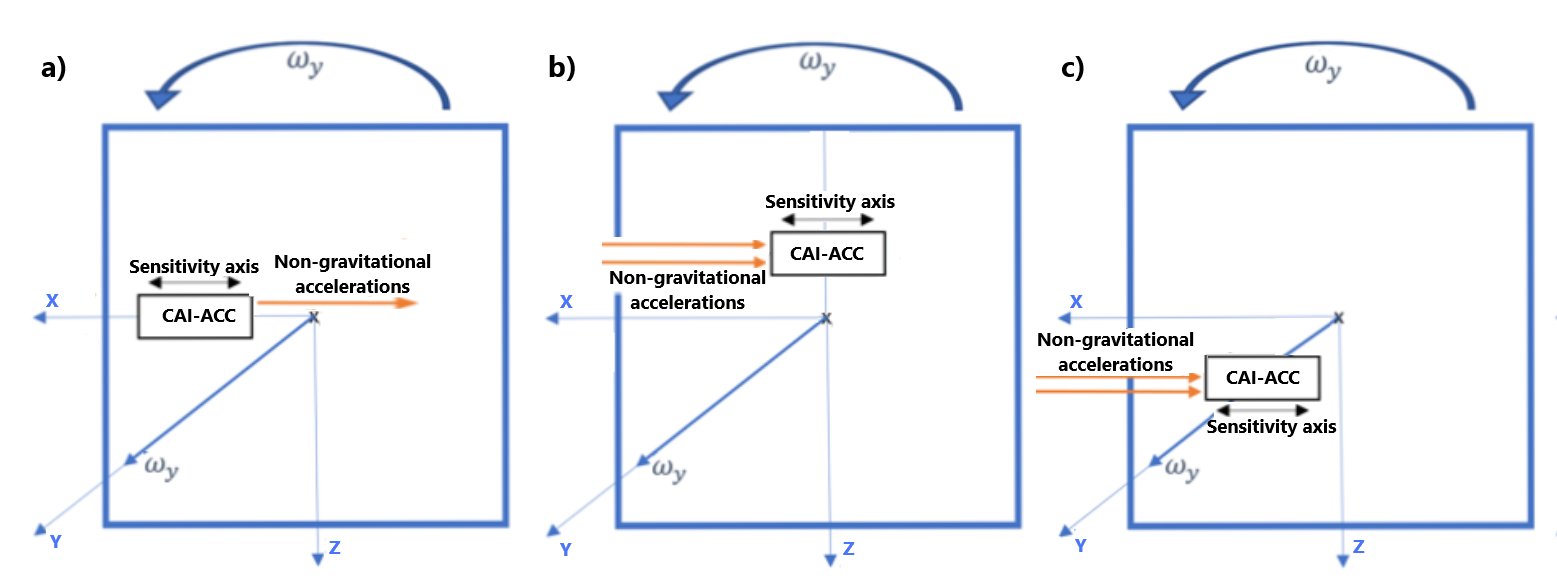}%
\caption{Positioning of the CAI accelerometer inside the satellite frame considering that the E-ACC is in the center of mass; a: The CAI accelerometer is placed on the along-track axis of the satellite with a displacement from the center of mass; b: The CAI accelerometer is placed on the radial axis of the satellite with a displacement from the center of mass; c: CAI accelerometer is placed on the cross-track axis of the satellite with a displacement from the center of mass; In all positions, the sensitivity axis of the CAI accelerometer is parallel with the along-track axis of the satellite. Only the along-track component of the non-gravitational accelerations, which is parallel with the sensitivity axis of the CAI accelerometer, is shown. }%
\label{fig:CAIpos}%
\end{figure*}

First, we assume that the CAI accelerometer is positioned in front of the E-ACC on the along-track axis of the satellite. This is a default position, actually considered in some studies \citet{Abrykosov2019,Zahzam2022}. We then study the three possible methods for the compensation of the rotation when the CAI accelerometer is placed in this position. We also try another possible configuration, in which we consider the CAI to be positioned on the cross-track axis of the satellite. There, the sensitivity axis would still be parallel to the along-track axis (see Fig.~\ref{fig:CAIpos}). Then, we investigate the impact of rotation when using active counter-rotating Raman mirrors. 
In this study, we do not consider a configuration similar to Fig.~\ref{fig:CAIpos}-b since it shows the worse result coming mostly due to the impact of the radial gravity gradient. 
%In the latter configuration, there would be no need for the technically challenging approach of rotating the whole CAI sensor and having counter-rotating mirrors would be sufficient to compensate for the rotation to an acceptable level. 

 Our simulations show that an error in the atomic cloud's initial positioning, combined with the uncompensated rotation rate, could result in an additional uncompensated centrifugal acceleration, causing noise in the measurements. This additional random noise would be one of the largest noise sources, especially in the case of rotation compensation using counter-rotating Raman mirrors.

Figure~\ref{fig:stateArt1} compares the impact of different assumptions for the rotation compensation for a state-of-the-art CAI accelerometer, indicating the importance of a proper rotation compensation technique. 
For the sake of simplicity, we will only consider the dominant rotation around the cross-track axis in the following discussion. However, the rotation rates around the other two axes, although much smaller, also result in additional terms for the Coriolis, centrifugal, and Euler accelerations and, thus, additional phase shifts. These additional error terms will be fully considered in the modeling of the CAI accelerometer noise.

%Better angular velocity compensation \\
% Less contrast loss due to Coriolis effect\\

\subsection{Sensitivity to the Atomic Temperatures}\label{sensiT0}

\begin{figure}%
\includegraphics[width=\columnwidth]{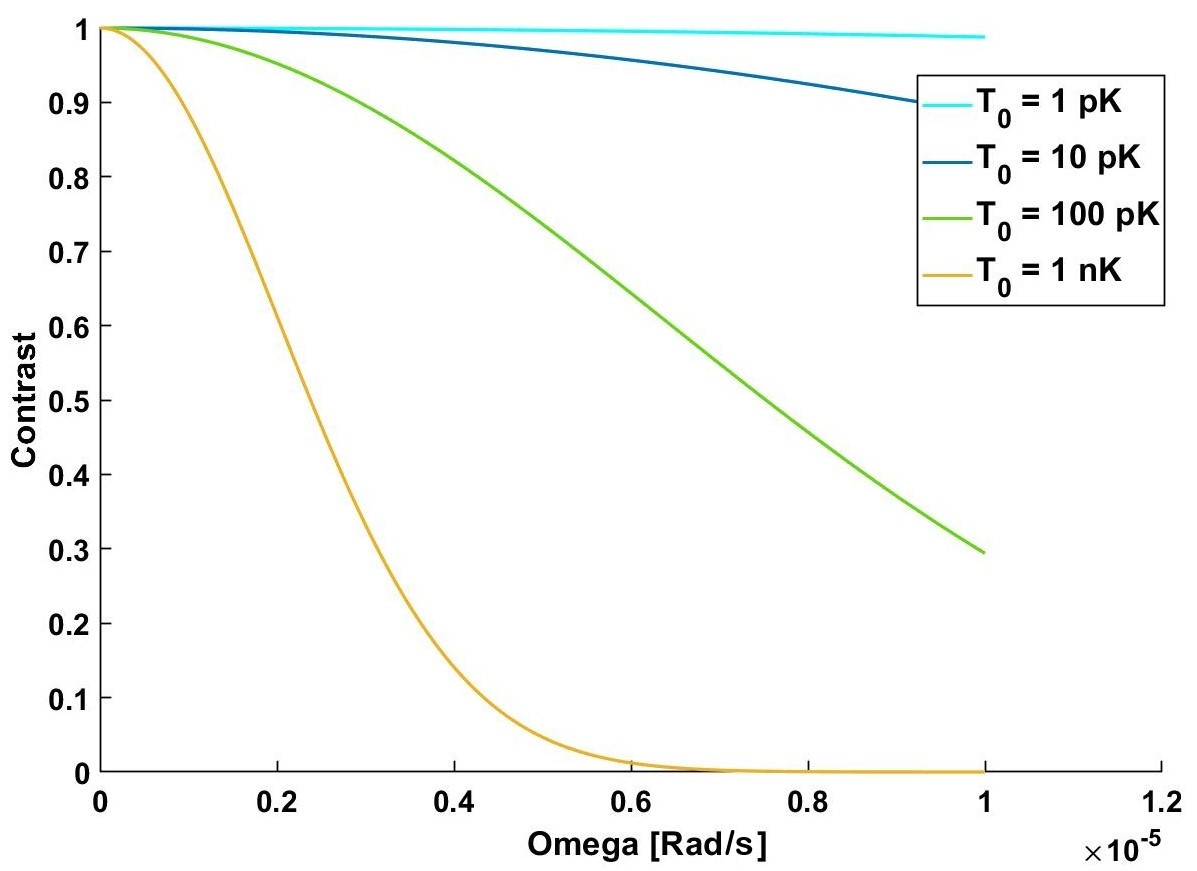}%
\caption{Contrast loss due to the Coriolis effect as a function of the uncompensated rotation rate of the satellite; Different colours represent different atomic temperatures} %
\label{fig:contLossCor}%
\end{figure}

The temperature of the atoms plays an important role in the performance of the CAI accelerometer measurements. Figure~\ref{fig:contLossCor} shows the contrast loss due to the Coriolis effect as a function of the rotation rate, for different temperatures and for an interferometer duration of $2T=\qty{10}{\s}$. For temperatures in the \unit{\nano \kelvin} range, uncompensated rotations of the order of a few \unit[per-mode=symbol]{\micro\radian\per\s} would lead to a significant loss of contrast. Reducing the temperature further down, into to the \unit{\pico \kelvin}, mitigates this effect.

\begin{figure}%
\includegraphics[width=\columnwidth]{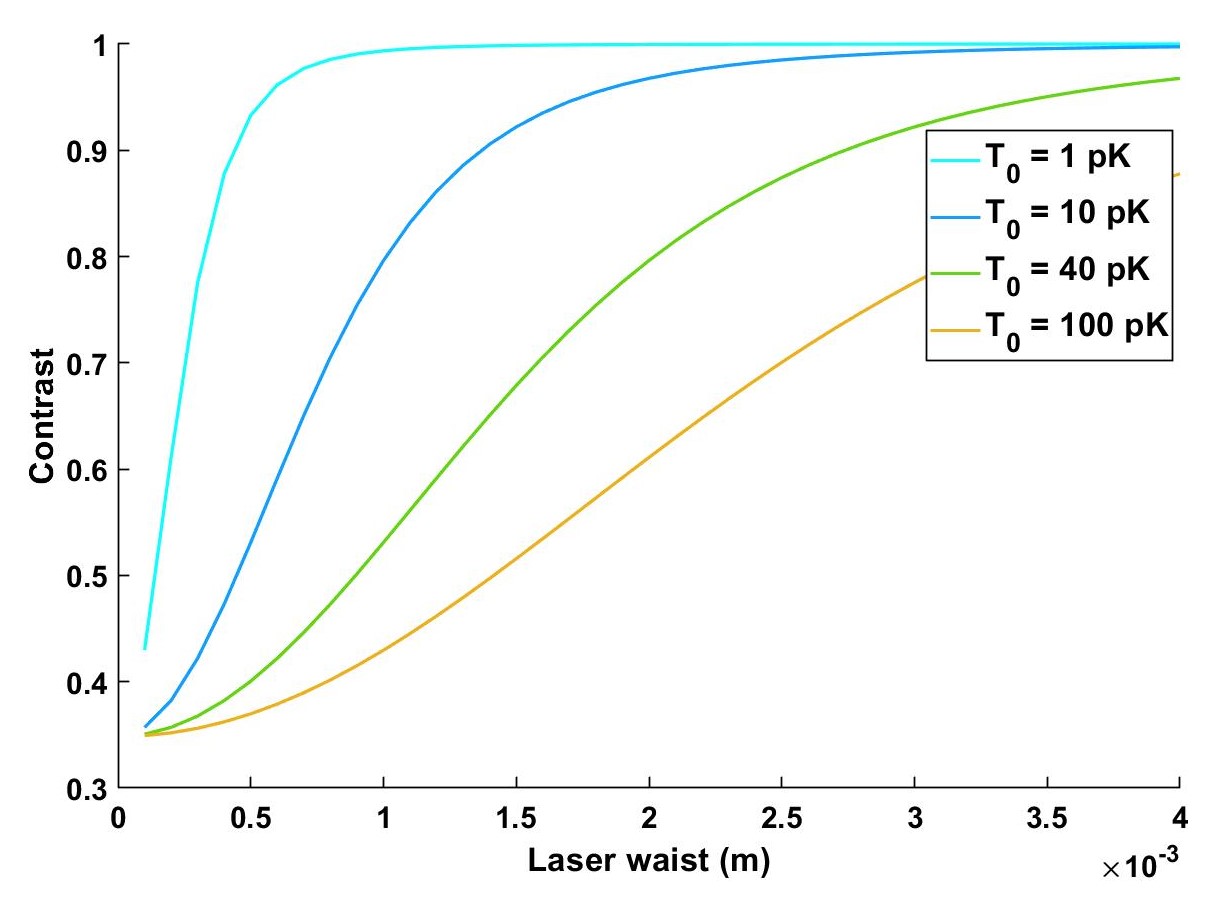}%
\caption{Contrast loss due to laser intensity inhomogeneity of an atom interferometer as a function of laser waist. Different colours represent different atomic temperatures.}%
\label{fig:contLossInho}%
\end{figure}

Figure~\ref{fig:contLossInho} illustrates the impact of laser intensity inhomogeneity by displaying the evolution of the contrast with laser waist for different temperatures. One can notice that increasing the laser waist is an efficient way to reduce contrast loss.  %However, it would not be helpful if the atomic temperatures were lower than the currently achievable atomic temperatures (\qty{40}{\pico \kelvin}). To achieve higher contrast, increasing the laser width should be combined with an improvement in the atomic temperatures.
 
  %Atomic temperatures also impact the wavefront aberration bias. With lower atomic temperature, the wavefront aberration bias also gets a smaller value. 

\subsection{Sensitivity to the Interrogation Time}\label{sensi2T}

 \begin{figure}
  \centering
  \includegraphics[width=\columnwidth]{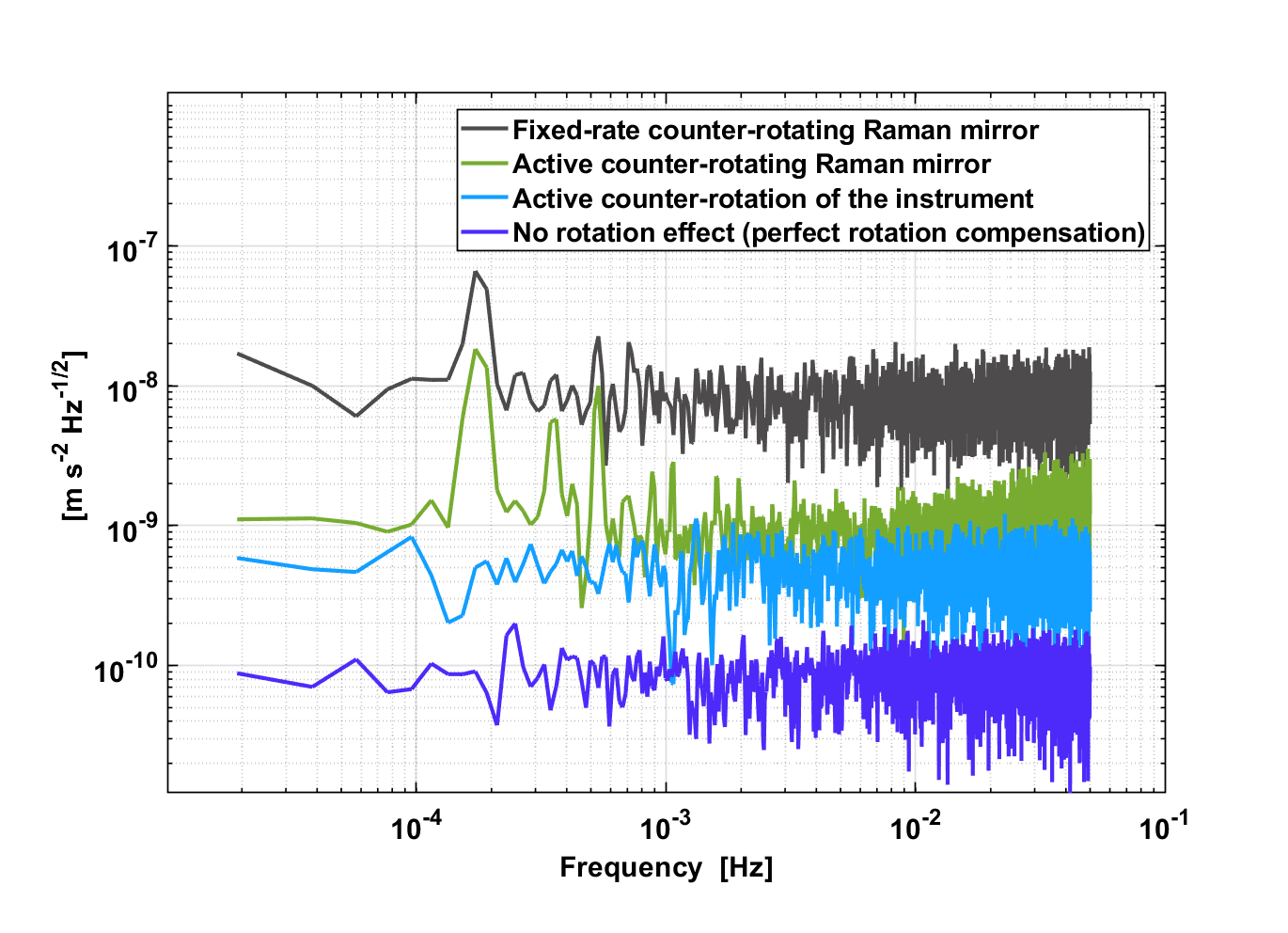}
  \caption{Amplitude spectral density for the CAI accelerometer measurement noise in the state-of-the-art scenario as defined in section~\ref{sec:stateArt}; Different colours represent different assumptions on the rotation compensation method of the CAI accelerometer. All curves are produced with the assumption that the CAI accelerometer is placed on the along-track axis of the satellite and in front of the E-ACC. In all scenarios, the sensitivity axis of the CAI accelerometer is assumed to be parallel with the along-track axis.}
  \label{fig:stateArt1}
\end{figure}

The interrogation time of the CAI accelerometer is a key parameter in the search for its optimal performance. On one hand, based on Eq.~\eqref{eq:CAIphase}, one expects a larger scale factor, and thus a higher measurement sensitivity. On the other hand, some components of the instrumental noise and errors, \eg, wavefront aberration and loss of contrast, get larger when increasing the interrogation time. 
Whether the performance would be better when increasing the interrogation time depends on the assumptions for the instrument parameters, rotation compensation, and satellite orbit. In practice, there will be an optimal interrogation time for each set of assumptions.

%Apart from these two effects, the variations of non-gravitational accelerations during the CAI interrogation time would also be higher with a longer interferometer duration.

%Figure~\ref{fig:sensiT} compares the total measurement noise of a state-of-the-art CAI accelerometer in different interrogation times. 
%Figure~\ref{fig:ultraSensi} shows the sensitivity of a far-future CAI accelerometer in long interrogation times.     

%Equations \ref{eq:}, and \ref{eq:} show the dependency of different noise sources on the CAI accelerometer interrogation time. With higher interrogation time, the wavefront aberration bias would have a higher value. ALso one would expect higher loss of contrast 
% Higher divergence of atomic cloud (higher contrast loss due to both cor. and laser inho. effects) 

%\subsection{Sensitivity to the Laser Waist}\label{sensiLaserWaist}

 %A higher laser waist would result in a lower loss of Contrast. 

%\subsection{Sensitivity to the Flatness of the Raman Mirror}\label{sensiT0}

%As it can be seen in figure~\ref{fig:stateNoiseTS}, the flatness of the Raman mirror is one of the limiting factors for the state-of-the-art sensors. By improving the flatness of the mirror, one would be able to lower the Wavefront aberrations bias and improve the total measurement noise. 

\subsection{Sensitivity to the Number of Atoms}\label{sensiN}

\begin{figure}
  \centering
  \includegraphics[width=\columnwidth]{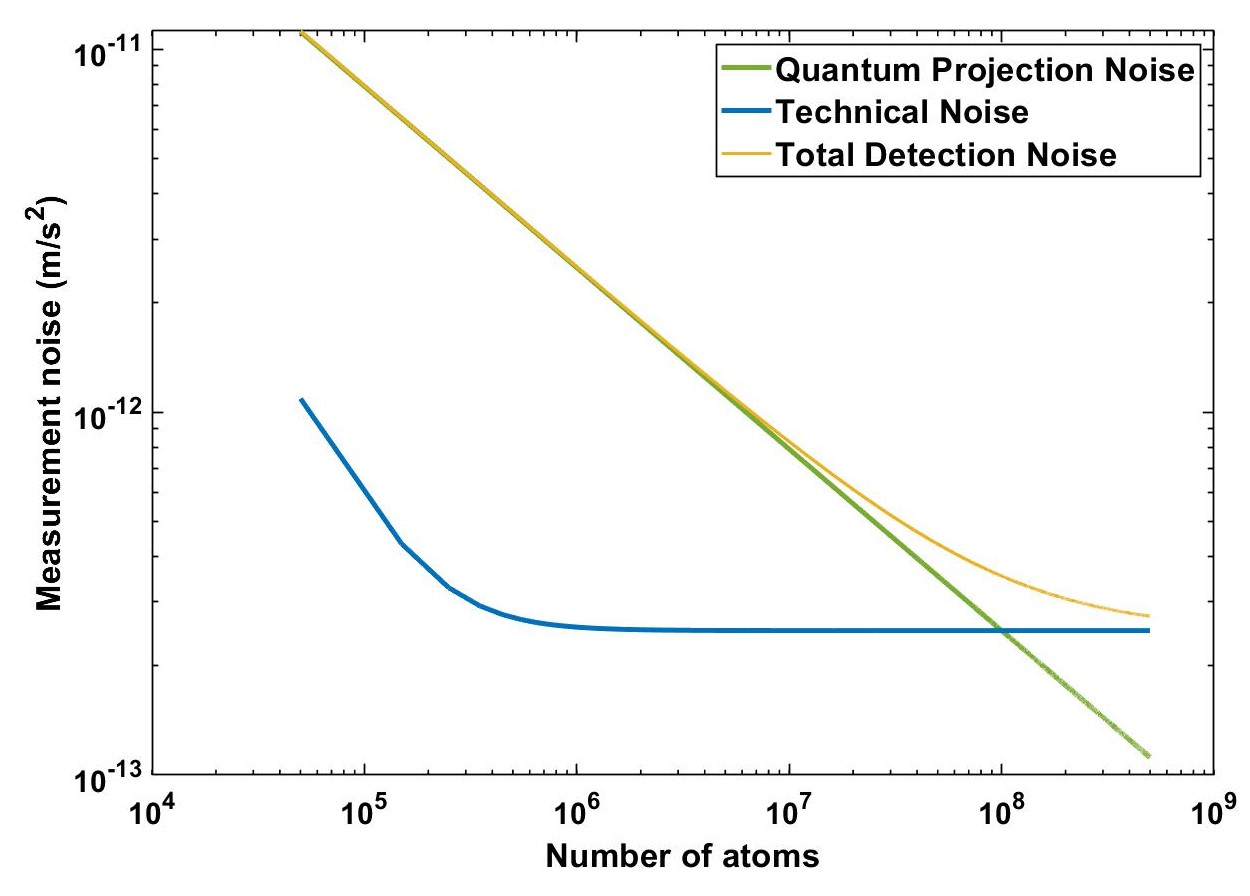}
  \caption{Sensitivity of the detection noise to the number of atoms; Green: Quantum projection noise; Blue: technical noise; Yellow: total detection noise. }
  \label{fig:QPN}
\end{figure}

Figure \ref{fig:QPN} shows the impact of increasing the number of atoms on the different components of the detection noise. One can see that a higher number of atoms can reduce both QPN and TN. With the current number of atoms in the CAI accelerometers based on the state-of-the-art technology (see table~\ref{tab:advancesTab}), the total detection noise is around \msqwert{1e-11}, which is more than one order lower than the other contributions of noise we have considered in the state-of-the-art scenario (see Figs. \ref{fig:stateArt2} and \ref{fig:noiseTS}). This noise level is also close to the other noise contributions in the near-future scenario. Therefore, we only consider a factor $2$ improvement in the number of atoms for the near-future scenario. %the target sensitivity for the near-future atom interferometry is within reach even with the current number of atoms. 

%From a technical point of view, it would be very challenging to increase the number of atoms for near-future missions, but it could be doable in far-future missions. Far-future atom interferometry will reach a level of sensitivity better than \msqwert{1e-11}, and from then on, it would be necessary to have a higher number of atoms in the atomic cloud. It is worth mentioning that after a certain point, the total noise reduction will be limited by a contribution of the technical noise which is independent of the number of atoms. 

 Far-future atom interferometry will reach a level of sensitivity better than \msqwert{1e-11}, and then, it would be necessary to improve the QPN in order to achieve better sensitivities. Improving the QPN limited sensitivity by increasing the number of atoms could be an option for far-future missions. However, the gain in sensitivity only scales as the square root of the atom number. Pioneering experiments on producing spin-squeezed states of atoms have shown a path toward preparing a state with large spin alignment and noise below the QPN level \citet{Anders2021,Greve2022}.  
 Therefore, a more suitable strategy for far-future missions could be implementing spin-squeezing techniques to overcome the standard quantum limit \citet{Gross2010,Hosten2016}.

%\begin{figure}
%  \centering
%  \includegraphics[width=\columnwidth]{figs/contrastLossCor.JPG}
%  \caption{Blue curve: LaserWaist = 6 mm; 
%Red curve: LaserWaist = 20 mm; green curve: Factor 5 improvement in the mirror flatness.}
%  \label{fig:contrastLossCor}
%\end{figure}
%\end{verbatim}

%\begin{figure}
%\centering
%\includegraphics[scale=0.5]{1.jpg}
%\caption{Write the figure caption here.}
%\label{fig:pendulum}
%\end{figure}

%\subsection{Compensation of the rotation}\label{subsecRot}

%\subsection{Equations}
%Conventionally, in mathematical equations, variables and
%anything that represents a value appear in italics. 
%All equations should be numbered for easy referencing. 
%The number should appear at the right margin.
%\begin{eqnarray}
%S'_{\mathrm{pg}} = \frac{S_{\mathrm{pg}} - \mathrm{min}(S_{\mathrm{pG}})}
 % {\mathrm{max}(S_{\mathrm{pG}} - \mathrm{min}(S_{\mathrm{pG}}))}
%\end{eqnarray}
%%In mathematical expressions 
%in running text "/" should be used
%for division (not a horizontal line).

\section{In-Orbit Performance Evaluation of CAI Accelerometry} \label{sec:orbitperf}

\subsection{State-of-the-art atom interferometry in space}\label{sec:stateArt}

We define the state-of-the-art CAI accelerometer in space as the quantum sensor that can be built with the currently available technology. The assumptions used for this scenario are briefly shown in table~\ref{tab:advancesTab}. 

\begin{figure}
  \centering
  \includegraphics[width=\columnwidth]{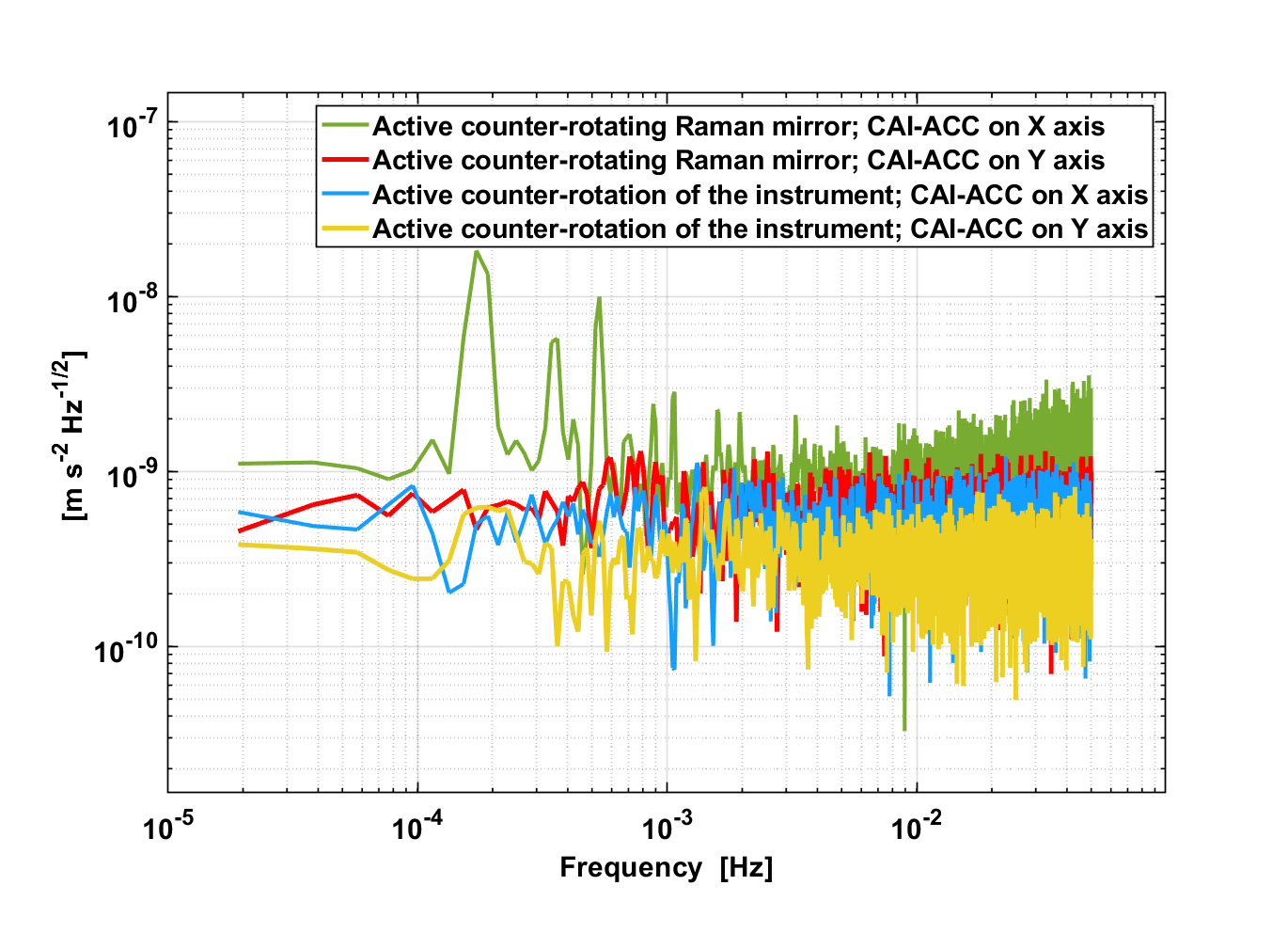}
  \caption{Amplitude spectral density for the CAI accelerometer measurement noise in the state-of-the-art scenario as defined in section~\ref{sec:stateArt}. Different colours represent different assumptions on the positioning and rotation compensation method of the CAI accelerometer. In all scenarios, the sensitivity axis of the CAI accelerometer is assumed to be parallel with the along-track axis.}
  \label{fig:stateArt2}
\end{figure}

Figure~\ref{fig:stateArt2} shows the noise of the state-of-the-art sensor with $\qty{10}{\s}$ interferometry duration in the frequency domain and compares the impact of different assumptions on the positioning and rotation compensation techniques on the noise level. Note that the counter-rotation of the whole quantum sensor would be technically challenging for the state-of-the-art scenario. We only show these plots to compare different rotation compensation methods. It is also important to take into account that in all these approaches, the sensitivity axis of the CAI accelerometer is still parallel to the along-track axis of the satellite and only the position of the CAI accelerometer inside the satellite is changed. The motivation for keeping the sensitivity axis along-track is that the non-gravitational acceleration in the along-track axis is the most critical component for a GRACE-like mission, i.e. the acceleration of the satellite along this direction directly impacts the determination of the gravity field. 

The green curve in Fig.~\ref{fig:stateArt2} shows the case where the CAI accelerometer is placed on the along-track axis of the satellite with a displacement from the center of mass (due to the positioning of E-ACC in the center of mass). According to Eq.~\eqref{eq:mirRot2}, not rotating the whole sensor, in this case, would result in a remaining term $x_0 {\Omega_y}^2$ which could cause a measurement bias. This term can, in principle, be mathematically corrected with proper knowledge of the rotational rates and the non-gravitational accelerations. 

Figure~\ref{fig:mathComp} shows an example of this bias before and after the mathematical correction. Our simulations show that the bias can be removed by two orders of magnitude, providing that a gyro with the accuracy presented in table~\ref{tab:advancesTab} exists. For all curves in Fig.~\ref{fig:stateArt1} and Fig. \ref{fig:stateArt2}, the mathematical correction is already applied.

%Figure~\ref{fig:stateArt2} is similar to figure~\ref{fig:stateArt}, with the difference that it 

Our simulations show that for the atom interferometry in space based on state-of-the-art technology, the optional interferometer duration $(2T)$ would be between $5$ to $\qty{10}{\s}$, with $2T=\qty{5}{\s}$ leading to slightly improved results. For near-future atom interferometry, $2T=\qty{10}{\s}$ would lead to the best performance. For the far-future scenario, it would be possible to increase the interrogation time even more. We demonstrate that an interferometer duration of $(2T)$ of $\qty{20}{\s}$ leads to considerably improved results. 

%, and~\ref{fig:stateNoiseTS}

%\subsection{State-of-the-art and Optimized State-of-the-art scenarios}\label{advances}

Improving quantum accelerometers beyond their current limits is a current challenge being tackled by a number of research groups worldwide. This gives way to the following prospective scenarios for future gravity missions with on-board quantum accelerometers.  

 %We first define an "optimized set-up" scenario with improvements of CAI accelerometers within reach in the next few years (<5 years). Here, we consider relatively modest technological improvements that could be planned for such a mission scenario. We will show that even such limited technological improvements could result in considerable improvements in the CAI accelerometers. We define 
 We first define a "near-future" scenario with improvements expected in the next 5-10 years. We then define a "far-future" scenario for expected improvements in the next 10-20 years.

% \begin{figure}
%   \centering
%   \includegraphics[width=\columnwidth]{figs/KK20.jpg}
%   \caption{Amplitude spectral density for the CAI accelerometer measurement noise in the optimized set-up as defined in section~\ref{sec:stateArt}; Different colours represent different assumptions on the positioning and rotation compensation method of the CAI accelerometer; In all scenarios, the sensitivity axis of the CAI accelerometer is assumed to be parallel with the along-track axis.}
%   \label{fig:OptiState}
% \end{figure}

\subsection{Future Advances in Atom Interferometry in Space}\label{futureCAI}

Intense research efforts are being made on scientific and engineering aspects to advance atom interferometry. All those aspects are important for atom interferometry. However, the question arises which aspects should be improved first or later to achieve maximum efficiency at each future period. The answer to this question would allow establishing a roadmap for future developments of quantum sensors for satellite gravity missions by improving the most impactful parameters. 

QPN acts as a natural noise limit for atom interferometry. For the near-future scenario, our strategy is to try to reduce other error sources down to the QPN level and find a path for future advances in atom interferometry, which optimizes the efforts. We would need a roadmap that both considers the impact on the performance of the CAI sensor and the technical limits. Table~\ref{tab:advancesTab} is created using this strategy and shows the assumptions for different mission scenarios.

\begin{figure*}[ht]
\includegraphics[width=\textwidth]{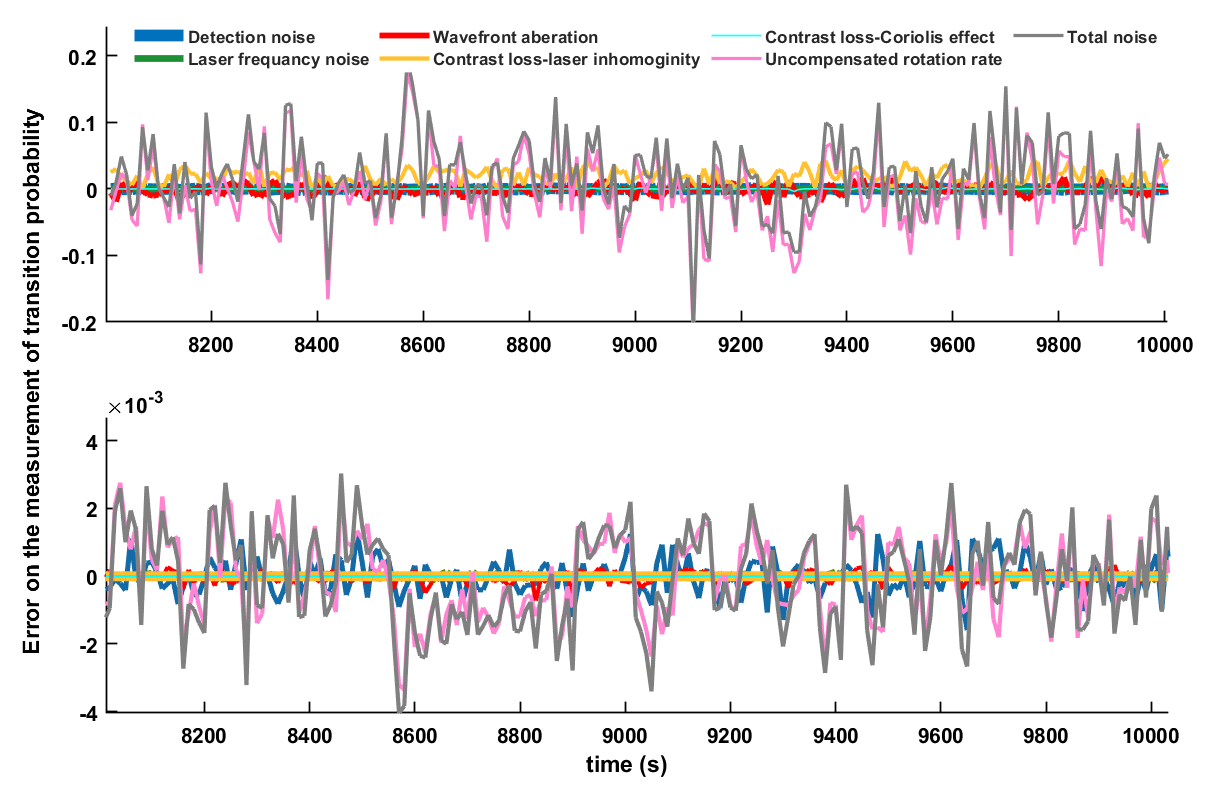}%
\caption{Simulation of the noise time series for a CAI accelerometer based on state-of-the-art technology (top plot), and for a near-future CAI accelerometer (bottom plot) for \qty{1000}{\s}. To compare the impact of different noise sources on the total noise, the components of the noise are plotted separately; To avoid the impact of initial fluctuations on the CAI measurements, we start the simulations of the CAI accelerometer after the stabilization of the satellite angular velocity (here, \qty{8000}{\s} after the start of the orbit propagation).}%
\label{fig:noiseTS}%
\end{figure*}

As shown in Fig.~\ref{fig:noiseTS}, 
the state-of-the-art scenario is limited by the contrast loss due to the laser intensity inhomogeneity and wavefront aberration noise. It is also considerably limited by the Coriolis and centrifugal accelerations caused by the uncompensated part of the rotation rate. QPN is a negligible noise source in this scenario. Going to the near-future scenario on the same figure, we notice that by the improvements suggested in table~\ref{tab:advancesTab}, all the noise sources have reduced to the level of QPN. 

%For the state-of-the-art atom interferometry scenario, we consider the measured wavefront of a high-quality commercial mirror, with a rated planeity of $\lambda/20$. 

%For the "Optimized-set-up" scenario, we consider increasing the laser waist to \SI{10}{\milli \meter} and improving the Raman mirror flatness by a factor of 5. Note that the impact of wavefront aberrations can also be reduced by improving the position stability of the atomic cloud. The comparison between figures ~\ref{fig:stateArt} and \ref{fig:OptiState} shows the benefit on the measurement sensitivity of these relatively modest improvements.
 \begin{figure}
  \centering
  \includegraphics[width=\columnwidth]{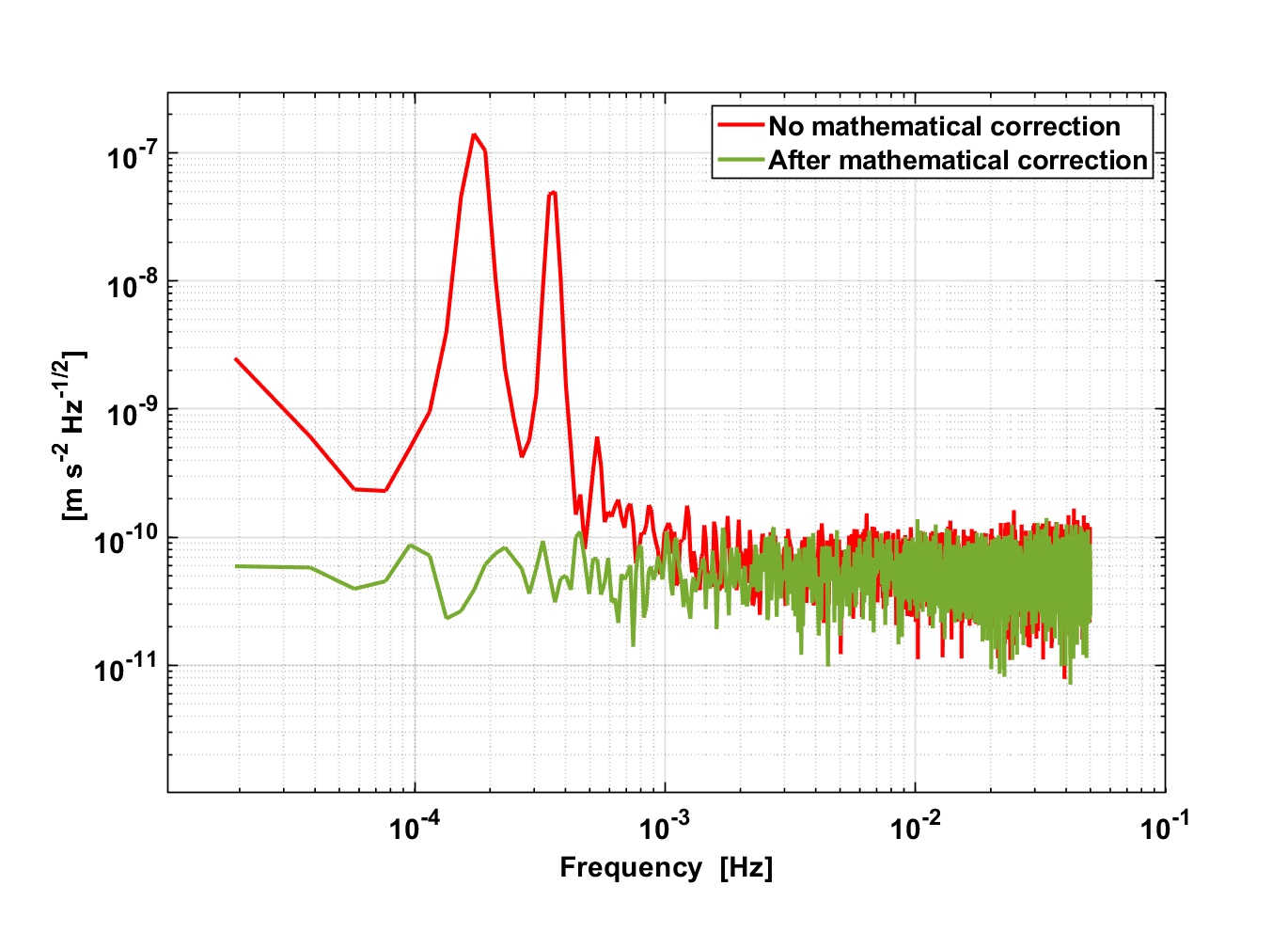}
  \caption{Amplitude spectral density for the CAI accelerometer measurement noise before and after the mathematical compensations of the low-frequency bias. The CAI accelerometer is positioned on the long-track axis of the satellite and in front of the E-ACC. The largest bias amplitude shows up at the orbital frequency.}
  \label{fig:mathComp}
\end{figure}

Figure~\ref{fig:stateArt1} shows the impact of different rotation compensation methods for the CAI accelerometer based on state-of-the-art technology. Without rotation compensation, the loss of contrast would limit the maximum interrogation time to a few hundred milliseconds only, so counter-rotation of the mirror is mandatory for performing acceleration measurements. With a fixed-rate counter-rotating Raman mirror, though the contrast would remain larger than 50\% at all times, the acceleration noise would remain limited by fluctuations of the residual Coriolis and centrifugal accelerations to the $10^{-8}~\textrm{m.s}^{-2}/\textrm{Hz}^{1/2}$ level. 

\begin{table*}[t!]
\centering
\caption{Assumptions for the CAI accelerometer on-board gravity missions } \label{tab:advancesTab}
\begin{tabular}{|l|c|c|c|}
\hline
 & CAI accelerometer & Near-future & Far-future \\
  Scenario           & based on  & CAI accelerometer & CAI accelerometer  \\
          & state-of-the-art technology & in space & in space \\
     \hline

\multirow{2}{*}{Expected time} & \multirow{2}{*}{now} & \multirow{2}{*}{next 5-10 years} & \multirow{2}{*}{next 10-20 years}   \\
    &   &  & \\
    \hline
\multirow{2}{*}{Laser Waist}  & \multirow{2}{*}{\qty{6}{\mm}}  & \multirow{2}{*}{\qty{20}{\mm}} & \multirow{2}{*}{\qty{40}{\mm}}     \\
    &   &  & \\
    %Laser Waist  &  $6 \times 10^6 m $       & $10 \times 10^6 m $        & $20 \times 10^6 m & $ $40 \times 10^6 m $        \\
\hline
\multirow{2}{*}{Atomic temperature ($ T_0 $)}  & \multirow{2}{*}{\qty{100e-12}{\kelvin}}  & \multirow{2}{*}{\qty{10e-12}{\kelvin}}  & \multirow{2}{*}{\qty{1e-12}{\kelvin}} \\
    &   &  & \\
    \hline
\multirow{2}{*}{Temperature stability}   & \multirow{2}{*}{\qty{4e-12}{\kelvin}}  & \multirow{2}{*}{\qty{1e-12}{\kelvin}}  & \multirow{2}{*}{\qty{0.5e-12}{\kelvin}} \\
    &   &  & \\
    \hline
\multirow{2}{*}{Number of atoms}    & \multirow{2}{*}{\num{5e+5}}  & \multirow{2}{*}{\num{1e+6}}  & \num{1e+7} or\\
    &   &  & \num{e+6} and 10 db squeezing\\
\hline
Technical noise    & \multirow{2}{*}{\num{1e-4}}  & \multirow{2}{*}{\num{1e-4} }    &  \multirow{2}{*}{\num{1e-5}}   \\
constant term ($\sigma_p (\infty)$)    &   &  & \\
\hline
\multirow{2}{*}{Rotation compensation} & \multirow{2}{*}{Counter-rotating mirror} & Counter-rotating mirror or & Counter-rotating mirror or \\

   &  &  counter-rotating CAI sensor & counter-rotating CAI sensor  \\
\hline
Initial positioning   & \multirow{2}{*}{\qty{1e-4}{\m}} & \multirow{2}{*}{\num{1e-4} to \qty{1e-5}{\m}} $^a$  &  \multirow{2}{*}{\num{1e-5} to \qty{1e-6}{\m}} $^b$\\
of the atomic cloud    &   &   & \\
\hline
Transversal velocity   & \multirow{2}{*}{\qty{100e-6}{\m\per\s}} & \multirow{2}{*}{\qty{20e-6}{\m\per\s}}   &  \multirow{2}{*}{\qty{5e-6}{\m\per\s}}  \\
of the atomic cloud    &   &   & \\
\hline
Noise of gyro  & \multirow{2}{*}{\qty{6.6e-7}{\radian\per\s}} & \multirow{2}{*}{\qty{6.6e-8}{\radian\per\s}} & \multirow{2}{*}{\qty{6.6e-9}{\radian\per\s}} \\
(white noise)  &        &       &       \\
\hline
Noise of E-ACC  & \multirow{2}{*}{\qty{1e-10}{\mps}} & \multirow{2}{*}{\qty{1e-11}{\mps}}  & \multirow{2}{*}{\qty{1e-12}{\mps}}  \\ 
(at higher frequencies)  &        &       &       \\
\hline
\multirow{2}{*}{Atomic flight time ($2T$)}  & \multirow{2}{*}{\num{5} to \qty{10}{\s}} & \multirow{2}{*}{\qty{10}{\s}}  & \multirow{2}{*}{\num{10} to \qty{20}{\s}}  \\ 
    &   &  & \\
\hline

\end{tabular}
\\
\footnotesize{ $^a$ In case of the counter-rotation of the whole CAI sensor, the accuracy of \qty{1e-4}{\m} would be enough; otherwise, we would need a positioning accuracy of \qty{1e-5}{\m}. $^b$ In case of the counter-rotation of the whole CAI sensor, the accuracy of \qty{1e-5}{\m} would be enough; otherwise, we would need a positioning accuracy of \qty{1e-6}{\m}}\\
\end{table*}

Better performances are met when actively compensating the rotations. When counter-rotating the mirror actively with the quantum accelerometer placed in front of the E-ACC on the along-track axis of the satellite, there remains a bias in the measurements though, fluctuating at the orbital frequency, related to the term discussed above in Equation \eqref{eq:mirRot2}. While this term could, in principle, be mathematically corrected for with proper knowledge of the rotational rates and the non-gravitational accelerations (see Fig.~\ref{fig:mathComp}), it could also be eliminated by either counter-rotating the whole CAI sensor inside the satellite (see the blue curve in Fig.~\ref{fig:stateArt2}), which would be technically very challenging, or more simply by positioning the CAI sensor on the cross-track axis of the satellite (the red curve in Fig.~\ref{fig:stateArt2}). Note that for all curves, the sensitivity axis of the CAI accelerometer is always assumed to be parallel with the along-track axis, and only the sensor's position is changed between the different scenarios.

While relatively small laser waists, in the range of \qtyrange{1.5}{6}{\milli \meter}, can be sufficient for ground-based measurements \citet{,,chiow2011102, merlet2014simple}, which motivated our choice of \qty{6}{\milli \meter} for the scenario based on state-of-the-art technology, we consider here a relatively large laser waist of \qty{20}{\milli \meter} for the near-future scenario. Reaching a Rabi frequency of \qty{16}{\kilo \hertz} at a Raman detuning of \qty{3.4}{\giga \hertz}, which optimizes the performance of the double-diffraction interferometer, demands laser powers of about \qty{100}{\milli \watt} and \qty{360}{\milli \watt} per beam, which matches with the typical powers achievable with current compact laser systems. Therefore, we consider this laser waist for near-future mission scenario. The required power scaling with the laser waist to the square, a waist of \qty{40}{\milli \meter} would lead to powers of about \qty{400}{\milli \watt} and \qty{1.44}{\watt}, which are more demanding in terms of technology. We have thus considered these laser parameters for the far-future mission scenario. 

As discussed, having an active rotation compensation method (\eg, active counter-rotating mirrors) is necessary in order to achieve a level of sensitivity close to the QPN level. For this, accurate gyros are needed to provide information about the satellite angular rate, cf. Table~\ref{tab:advancesTab}. For the state-of-the-art mission, we assume gyros similar to the gyro of GRACE-FO satellites (see Table~\ref{tab:advancesTab}). However, for the near-future and far-future mission scenarios, we assume gyros with noise performances of one and two orders of magnitude better than the state-of-the-art ones.

\begin{figure}
  \centering
  \includegraphics[width=\columnwidth]{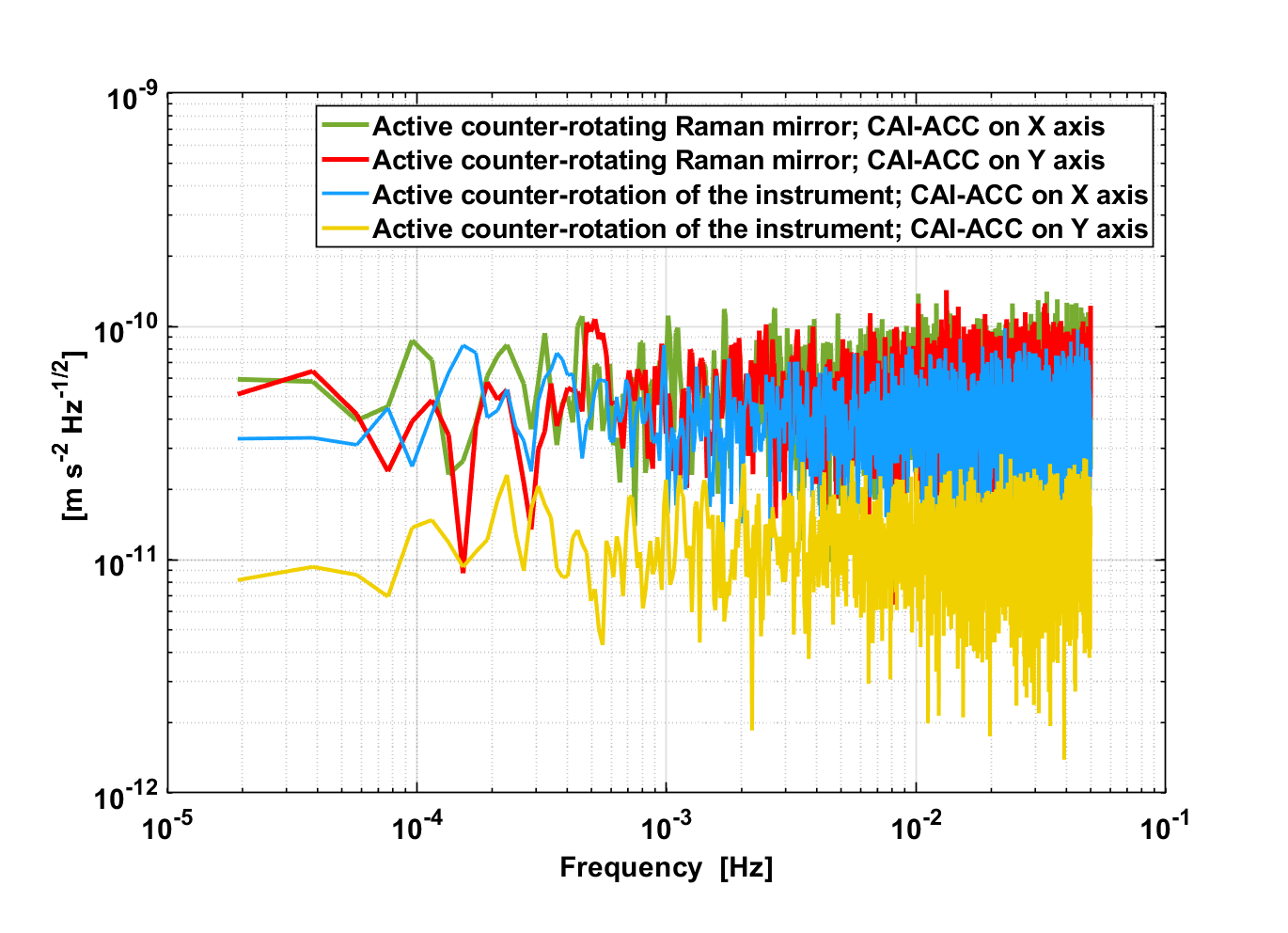}
  \caption{Amplitude spectral density for the CAI accelerometer measurement noise in the near-future scenario. Different colours represent different assumptions on the positioning and the rotation compensation method of the sensor. }
  \label{fig:nearFut}
\end{figure}

As discussed in section~\ref{sec:Intro}, having in view a hybrid accelerometer configuration, we consider the integration of an electrostatic accelerometer in parallel with the CAI accelerometer. To fully benefit from such a hybrid configuration, the sensitivity floor of the E-ACC at higher frequencies should be at the same level as the CAI accelerometer. Therefore, for the near-future and far-future mission scenarios, we assume an E-ACC one order and two orders of magnitude more sensitive than the current state-of-the-art E-ACCs. 

The lowest atomic temperature demonstrated by now is around \qty{40}{\pico \kelvin}, achieved with the Delta kick method \citet{deppner2021, xie2022ground}. We thus assume this temperature for the state-of-the-art scenario. Based on the ongoing advances in this field, an atomic temperature of \qty{10}{\pico \kelvin} is considered for near-future missions and an atomic temperature of \qty{1}{\pico \kelvin} for far-future missions.

Based on the assumptions we consider for the near-future mission scenario, all the instrumental and environmental noise components reach a noise level close to QPN. Beyond this, the only way to improve the CAI accelerometer would be to reduce the QPN by increasing the number of atoms or implementing quantum metrology protocols to bring down the detection noise below the standard quantum limit, implementing quantum correlations in the atomic source via spin-squeezing for instance. Therefore, for the far-future scenario, we consider an order of magnitude improvement in the detection noise variance, which we believe is within reach.

\begin{figure}
  \centering
  \includegraphics[width=\columnwidth]{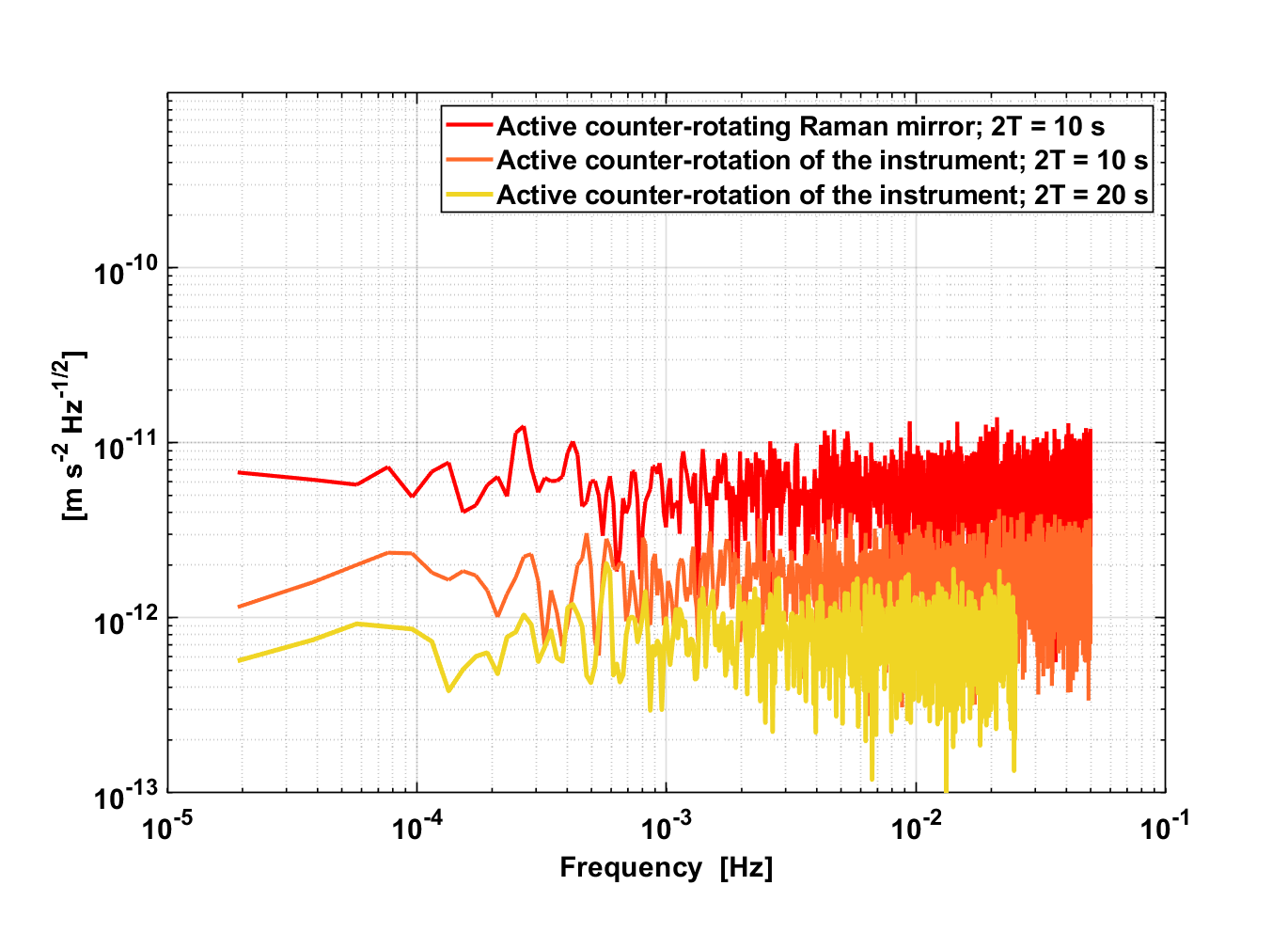}
  \caption{Amplitude spectral density for the CAI accelerometer measurement noise in the far-future scenario; Different colours represent different assumptions on the positioning and rotation compensation method of the CAI accelerometer. }
  \label{fig:farFut}
\end{figure}

We finally perform several simulation studies based on the modeling explained in section~\ref{sec:modeling} and based on the different sets of assumptions that are shown in~Table~\ref{tab:advancesTab}. Figures~\ref{fig:nearFut} and~\ref{fig:farFut} show the estimated measurement noise of the near-future and far-future quantum accelerometers with different assumptions for the sensor.

While the difference between rotation compensation using counter-rotating Raman mirrors and using the counter-rotation of the whole quantum sensor is small for state-of-the-art missions (see Fig.~\ref{fig:stateArt2}), these techniques could result in a difference of about a factor of 3 or 4 in the noise level of the near and far future scenarios (see Fig.~\ref{fig:nearFut} and Fig. \ref{fig:farFut}). A large part of this difference is a direct result of an error in the initial positioning of the atomic cloud, which causes a large uncompensated centrifugal acceleration for the case we only rotate the Raman mirror (see Eq. \eqref{eq:mirRot2}).

In the near future, it would be enough if the Raman mirror is rotated against the satellite's main rotation around the cross-track axis. However, we found that for the far-future scenario, to compensate for the rotation to the required level, it would be necessary to also rotate the Raman mirror against the smaller rotation rate of the satellite around the radial axis. 

While the optimal interferometry duration for the near-future scenario is around $\qty{10}{\s}$, for the far-future scenario, increasing the interferometry duration beyond $\qty{10}{\s}$ would be possible. The optimal interferometry duration for this scenario is around $\qty{20}{\s}$, which results in a noise level lower than \msqwert{1e-12} in case of the counter-rotation of the whole sensor. For the case of counter-rotation of the Raman mirror, increasing the interferometry duration beyond $\qty{10}{\s}$ would not improve the solution. 

\begin{figure}
  \centering
  \includegraphics[width=\columnwidth]{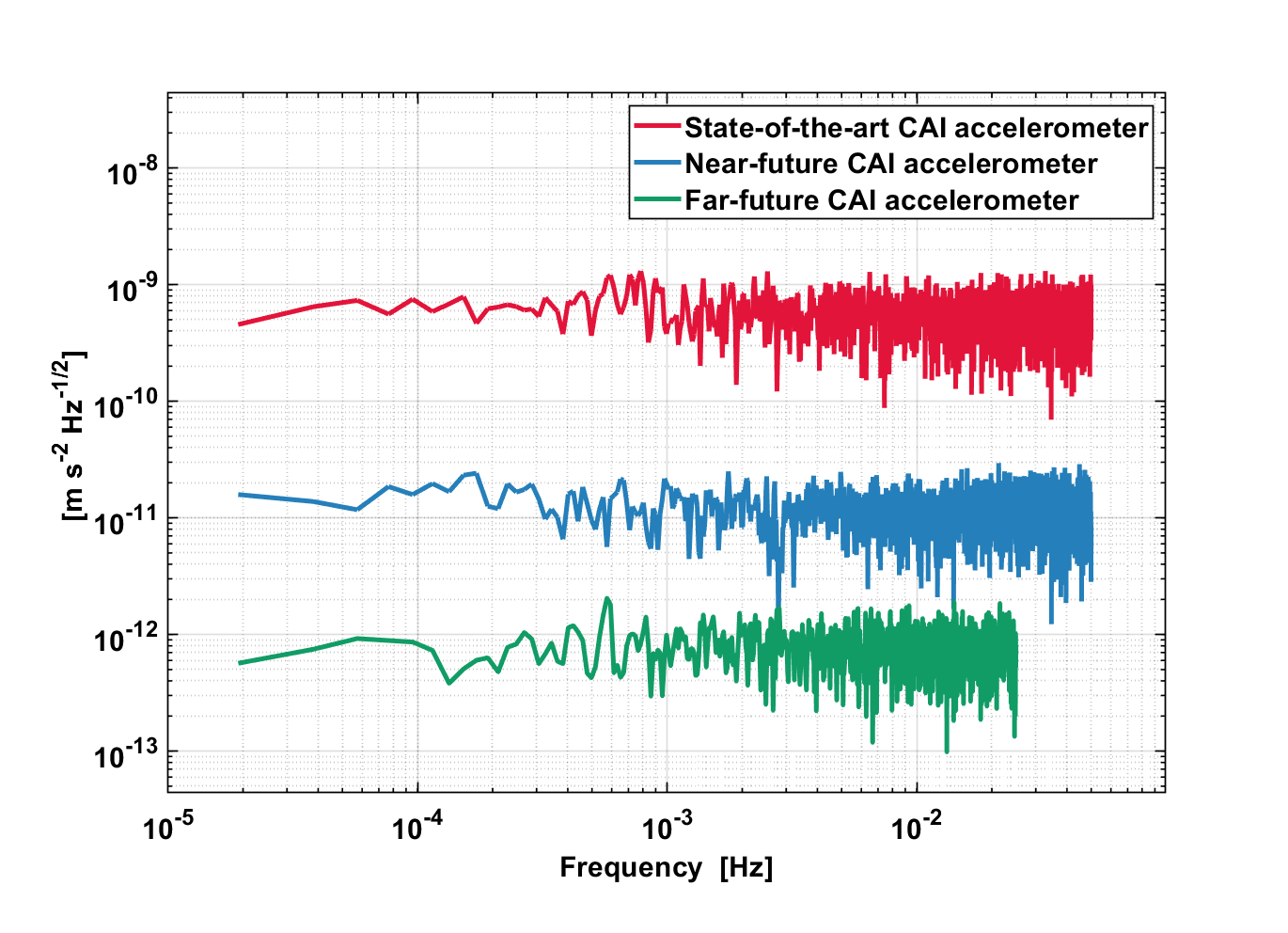}
  \caption{Comparison of the amplitude spectral densities for the CAI accelerometer measurement noises in the state-of-the-art, near future (in 5-10 years) and far-future (in 10-20 years) scenarios; For all scenarios active counter-rotating Raman mirrors are assumed and the CAI accelerometer is positioned on the cross-track axis of the satellite.}
  \label{fig:ASDall}
\end{figure}

Figure~\ref{fig:ASDall} compares the amplitude spectral density of the CAI accelerometer measurement noise for the different scenarios in one plot considering the use of an active counter-rotating Raman mirror and the positioning of the CAI accelerometer on the cross-track axis of the satellite (see Fig.~\ref{fig:CAIpos}).

%considering the improvements mentioned in table~\ref{tab:advancestab}, we can achieve the following performance for the CAI accelerometer for near-future satellite gravity missions. 

%We provide this roadmap by studying the sensitivity plots to learn the most impactful parameters and also by studying the estimated noise components and comparing them to the natural QPN noise level. Finally, the technical possibilities are applied and a roadmap is 

%We can use this approach to find the most optimized. 

%Considering the current technology, there are a few improvements which could be made in the next 5-10 years...

%Higher laser weist
%The flatness of the mirror
%Quantum projection noise

%\subsection{And a table?}
%Just replace the text/values in the template table below 
%with your own. You can change the number of 
%lines/rows as necessary.

%\subsection{Optimized-State-of-the-art atom interferometry in space}\label{advances}

%\subsection{Near-future atom interferometry in space}\label{advances}

%\subsection{Far-future atom interferometry in space}\label{advances}

%\subsection{Ultra-sensitive atom interferometry in space}\label{advances}

\subsection{Ultra-Sensitive Atom Interferometry in Space}

Since future space missions would require even more sensitive absolute accelerometers, we discuss in this section the possibility of pushing further down the sensitivity of quantum accelerometers. As discussed in section~\ref{signalModel}, the sensitivity of a quantum accelerometer increases with a longer interrogation time, and being in space would allow to benefit from this. State-of-the-art atom interferometry, still under development for space applications, is considering an interrogation time in the order of a couple of seconds~\citet{Leveque2022, Beaufils2023, Zahzam2022}. In this study, we have considered the state-of-the-art and near-future quantum accelerometers to have interferometry duration between $2T =\num{5}$ to $\qty{10}{\s}$ and for the far-future quantum sensors, we show that we can reach an interferometry duration of $2T = \qty{20}{\s}$. But, with the advances that are expected in this field, even longer interrogation times will be possible.

\begin{figure}
  \centering
  \includegraphics[width=\columnwidth]{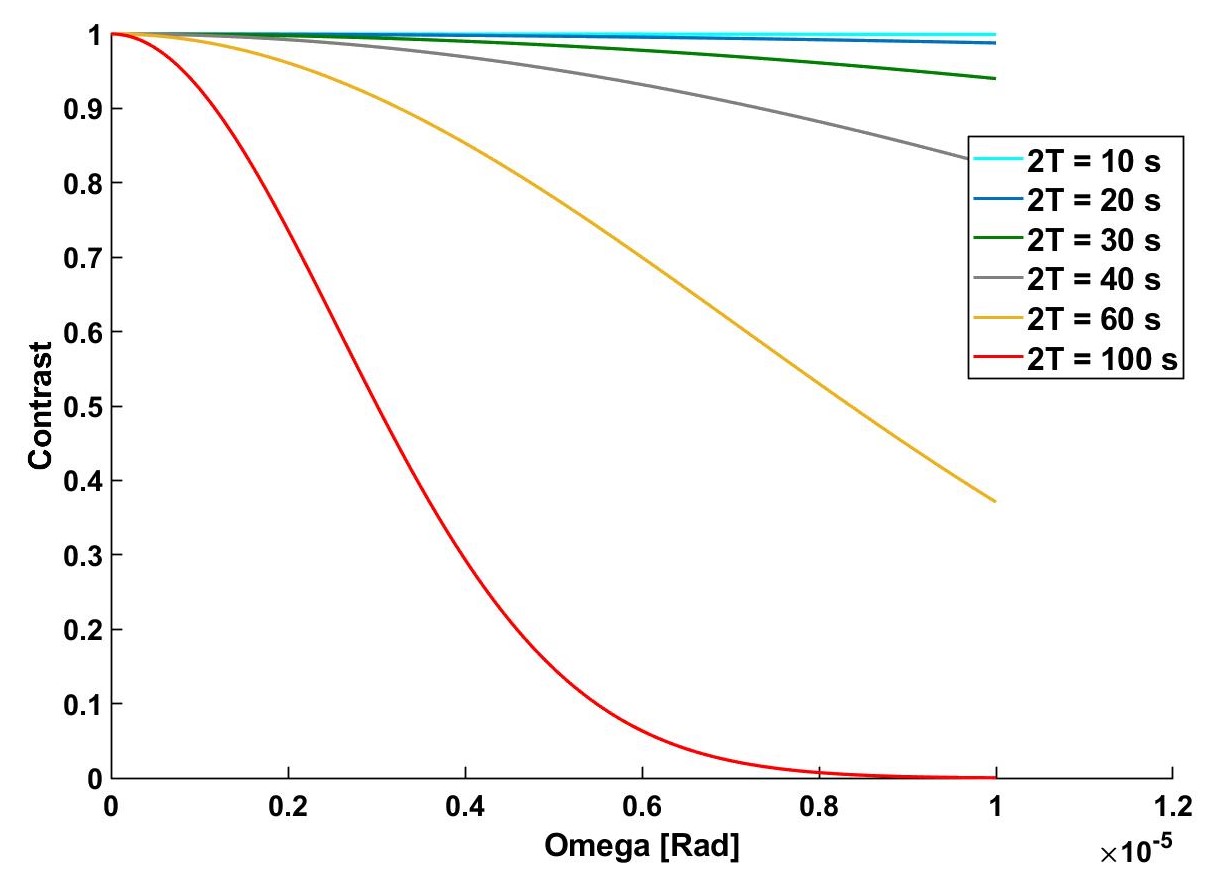}
  \caption{Long-interrogation-time atom interferometry in space; In all cases, assumptions of the far-future atom interferometry in space (see Table~\ref{tab:advancesTab}) are applied together with a longer interrogation time}
  \label{fig:contLossLongT}
\end{figure}

To have a quantum accelerometer with longer interrogation times, one would need to consider that apart from getting larger signals, some components of the error (e.g. contrast loss) would also get larger magnitudes. Really benefiting from longer free flight times will demand to carefully limit these error sources. Figure~\ref{fig:contLossLongT} displays the loss of contrast as a function of the uncompensated part of the rotation for different interrogation times, showing that the drop in contrast gets much higher with longer interrogation times. Keeping a low contrast loss with high interrogation time will thus demand extensive compensation of all rotational effects. Longer interrogation times also lead to a larger physical package size and will demand wider laser beams and higher powers. They will also need improved mirrors in terms of flatness to avoid the wavefront aberration noise from getting large values. 

\begin{figure}
  \centering
  \includegraphics[width=\columnwidth]{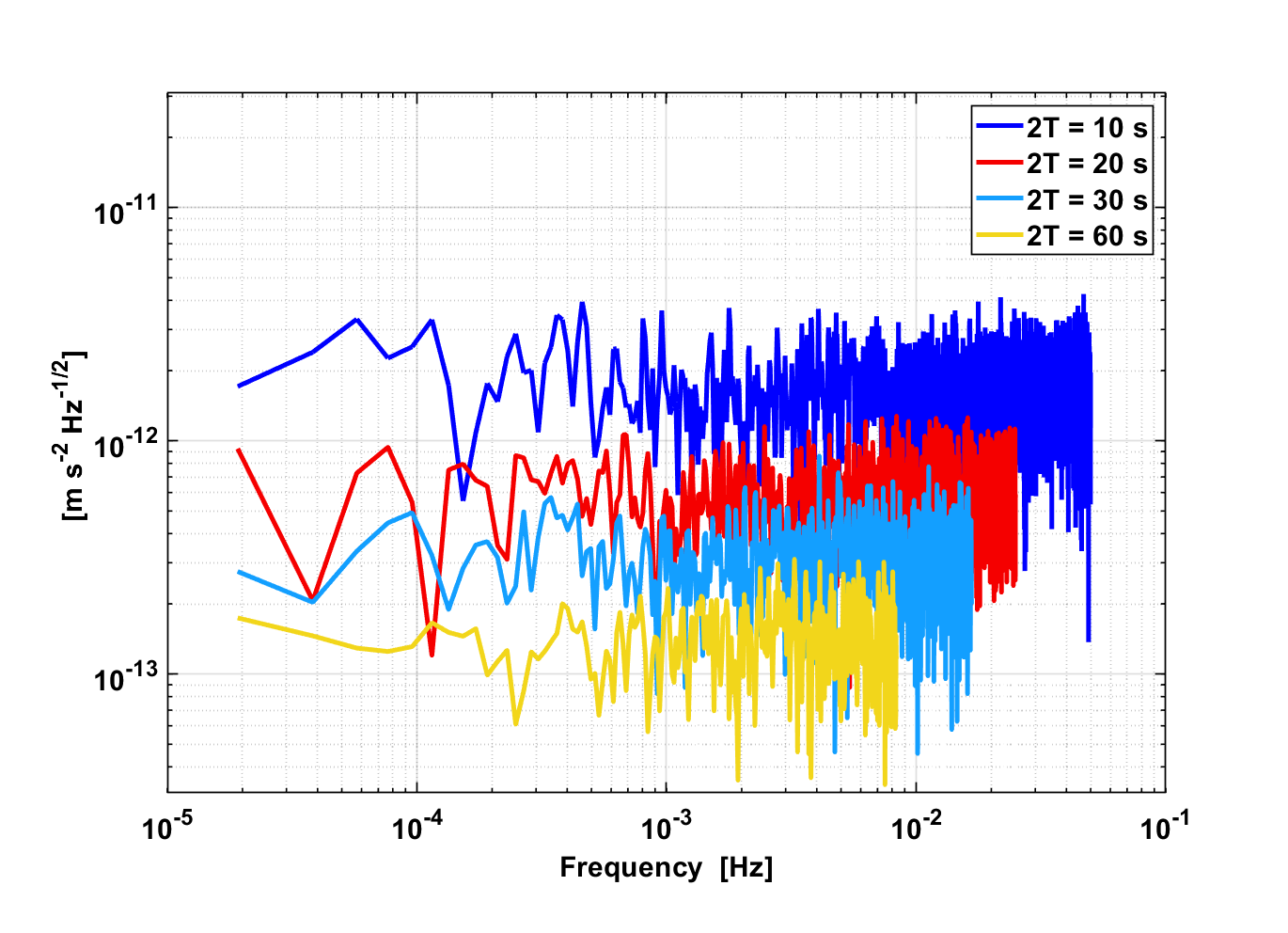}
  \caption{Long-interrogation-time atom interferometry in space. In all cases, assumptions of the far-future atom interferometry are applied with the difference of having longer interrogation times and a comparable E-ACC in terms of accuracy in higher frequencies. With higher interrogation time, the sensitivity of the instrument will increase according to Eq.~\eqref{eq:CAIphase}. However, there is also an increase in several error sources (e.g. contrast loss due to the Coriolis effect). The total noise level is calculated considering the combination of all these effects.}
  \label{fig:ultraSensi}
\end{figure}

Figure~\ref{fig:ultraSensi} compares the total measurement noise of CAI accelerometers with different interrogation times in the frequency domain, assuming that all rotational effects are fully compensated. Sensitivity levels down to \msqwert{1e-13} could be reached, extrapolating interferometer duration up to $\qty{60}{\s}$. Increasing the interferometry duration beyond this would result in a drop in the sensitivity.

\section{Conclusions}

Quantum accelerometers are foreseen for future satellite gravity missions. In this study, we develop a comprehensive in-orbit performance model for a quantum accelerometer onboard future gravity missions, in which we study the impact of various sources of errors on the stability of cold atom interferometers. We investigate their performance under different assumptions about the positioning and rotation compensation method, and we conclude that without an active rotation compensation method which employs an actuator to create a counter-rotation based on the gyro data at each instant in time, the errors will be so large as to prohibit from benefiting from the instrument. 

We also show that in the scenario where the CAI accelerometer is placed on the along-track axis, and the rotation is compensated by using an active counter-rotating Raman mirror, the remaining part of the uncompensated rotation will cause a relatively large bias in the CAI accelerometer measurements, compromising the ability of this instrument to act as an absolute inertial instrument. In this scenario, a mathematical calculation and correction of the bias based on the gyro data would reduce the bias by about two orders of magnitude. 

We found that the highest sensitivity is achieved by positioning the E-ACC in the center of mass and the CAI accelerometer beside the E-ACC on the cross-track axis of the satellite. In this scenario, having active counter-rotating Raman mirrors with the rest of the instrument attached to the satellite frame would be sufficient to compensate for the rotational effect. However, for the near and far-future scenarios, counter-rotating the whole sensor would result in factor 3 to 4 improvement in the noise level.

%We also show that in the scenario where the CAI accelerometer is placed on the along-track axis, the full compensation of the rotational effects is possible only with the whole instrument actively counter-rotating against the satellite rotation rate, which would be challenging from the technical point of view. In the scenario where the CAI accelerometer is located on the along-track axis and the rotation is compensated by using active counter-rotating Raman mirrors, the remaining part of the uncompensated rotation will cause a relatively large bias in the CAI accelerometer measurements, compromising the ability of this instrument to act as an absolute inertial instrument. 

We also discuss current and future advances expected for space-based atom interferometry and investigate their impact on the performance of the CAI accelerometers in different scenarios. First, we consider atom interferometry based on state-of-the-art technology, and we show that a stability level of \msqwert{5e-10} can be achieved. 
The second scenario is a near-future scenario, where we consider an improved quantum accelerometer based on the technological progress expected in the next 5 to 10 years. An expected sensitivity level of \msqwert{1e-11} to \msqwert{5e-11} is estimated depending on the rotation compensation method. 

Then, we consider a far-future scenario and select it as the most efficient quantum accelerometer achievable in the next 10 to 20 years, and we demonstrate that sensitivity in the order of \msqwert{1e-12} could be achieved in this scenario. The major challenges in achieving this sensitivity would be, first, an accurate rotation compensation by a more accurate initialization of the atomic cloud and by employing improved gyros, E-ACC, and Raman mirror actuators with respect to state-of-the-art technology. The other challenge would be an increase in the number of atoms in the atomic cloud to $10^7$. This would be necessary to reduce the quantum projection noise, which turns out to be critical at this level of stability. An alternative approach would be to apply a squeezing technique to reduce the QPN by the same magnitude. 

Based on our simulation results, we provide a roadmap for advances in atom interferometry that would maximize the performance of future CAI accelerometers, taking into account possible improvements as well as technical challenges. We conclude that for each set of assumptions for the quantum sensor, there is an optimal interrogation time. The optimal interferometry duration are respectively \qty{5}{\s}, \qty{10}{\s} and \qty{20}{\s} for the state-of-the-art, near-future and far-future scenarios. 
Finally, we discuss the possibility and challenges of having ultra-sensitive atom interferometry for future space missions by considering longer interrogation times in space. Apart from its applications for future satellite gravimetry, this ultra-sensitive quantum sensor could be very attractive for space-based experiments of fundamental physics.

%Citations in the text can be made using\\[6pt]
%\verb+\citet{NewmanGirvan2004}+\\[6pt]
%for citation in running text like in 
%\citet{NewmanGirvan2004} or using\\[6pt]
%\verb+\citet{Vehlowetal2013,NewmanGirvan2004}+\\[6pt]
%for citation within parentheses like in 
%\citet{Vehlowetal2013,NewmanGirvan2004}.

%Please use the actual \verb+\cite+ command in the text.
%Also, please double-check the \verb+\citet+ command.

%\section{Reference style}
%You can include the references in the main text file in \LaTeX
%format. Alternately, you can include a separate bibliography
%file (refs.bib in this example) and run the following set of 
%commands:

\section*{Acknowledgments}
This work is supported by the Deutsche Forschungsgemeinschaft (DFG, German Research Foundation) Collaborative Research Center 1464 “TerraQ” – 434617780 and Germany’s Excellence Strategy – EXC-2123 “QuantumFrontiers” – 390837967, and by the European Union’s Horizon 2020 research and innovation programme under grant agreement No 101081775 (CARIOQA-PMP project). This study is also partially supported by SpaceQNav project. 
QB and FP acknowledge the support from a government grant managed by the Agence Nationale de la Recherche under the Plan France 2030 with the reference “ANR-22-PETQ-0005”.
B.T. acknowledges support from the Federal Ministry for Economic Affairs and Energy (BMWi), Project 50RK1957. A.K. acknowledges support by Deutsches Zentrum für Luft- und Raumfahrt e.V. (DLR) for the project Q-BAGS.

%% Bibliography
%% Author year style
%\bibliographystyle{jasr-model5-names}

%\biboptions{authoryear}
\bibliographystyle{plainnat}

\bibliography{refs2.bib}

%\bibliographystyle{abbrvnat}
% \section*{Acknowledgments}
% We thank you for reading these guidelines.

 %\bibliographystyle{authoryear}  
% \bibliography{references.bib}
\end{document}